\documentclass[manuscript,onefignum,onetabnum]{siamart190516}

\usepackage{lipsum}
\usepackage{amsfonts}
\usepackage{graphicx}
\usepackage{epstopdf}
\usepackage[margin=2.5cm]{geometry}
\usepackage{subfig} 
\usepackage{bm}
\usepackage{bbm}
\usepackage{amsmath}
\usepackage{multirow}
\usepackage{pdflscape}
\usepackage{mathtools}
\usepackage{tabularx}
\usepackage{overpic}
\usepackage{placeins}
\usepackage{caption}
\usepackage{amssymb}
\usepackage{cleveref}
\usepackage{lineno}
\usepackage{algorithm}
\usepackage[noend]{algpseudocode}
\usepackage{etoolbox}
\usepackage{tikz}
\usetikzlibrary{arrows.meta}
\tikzset{%
	>={Latex[width=2mm,length=2mm]},
	base/.style = {rectangle, rounded corners, draw=black,
		minimum width=2cm, minimum height=1.25cm,
		text centered,font=\rmfamily\footnotesize},
	output/.style = {base, fill=black!15},
	input/.style = {base, fill=green!15},
	truth/.style = {base, fill=black!35},
	process/.style = {base, fill=white}
}

\crefname{equation}{Eq.}{Eqs.}
\crefname{figure}{Fig.}{Figs.}
\crefname{tabular}{Tab.}{Tabs.}
\crefname{section}{Section}{Sections}
\crefname{subsection}{Subsection}{Subsections}
\crefname{algorithm}{Algorithm}{Algorithms}

\makeatletter
\newcommand*{\algrule}[1][\algorithmicindent]{%
	\makebox[#1][l]{%
		\hspace*{.2em}
		\vrule height .75\baselineskip depth .25\baselineskip
	}
}

\newcount\ALG@printindent@tempcnta
\def\ALG@printindent{%
	\ifnum \theALG@nested>0
	\ifx\ALG@text\ALG@x@notext
	\else
	\unskip
	\ALG@printindent@tempcnta=1
	\loop
	\algrule[\csname ALG@ind@\the\ALG@printindent@tempcnta\endcsname]%
	\advance \ALG@printindent@tempcnta 1
	\ifnum \ALG@printindent@tempcnta<\numexpr\theALG@nested+1\relax
	\repeat
	\fi
	\fi
}
\patchcmd{\ALG@doentity}{\noindent\hskip\ALG@tlm}{\ALG@printindent}{}{\errmessage{failed to patch}}
\patchcmd{\ALG@doentity}{\item[]\nointerlineskip}{}{}{} 
\makeatother

\makeatletter
\def\algbackskip{\hskip-\ALG@thistlm}
\makeatother

\ifpdf
  \DeclareGraphicsExtensions{.eps,.pdf,.png,.jpg}
\else
  \DeclareGraphicsExtensions{.eps}
\fi

\newsiamremark{remark}{Remark}
\newsiamremark{hypothesis}{Hypothesis}
\crefname{hypothesis}{Hypothesis}{Hypotheses}
\newsiamthm{claim}{Claim}

\DeclareMathOperator*{\argmin}{arg\,min}
\DeclareMathOperator*{\argmax}{arg\,max}

\headers{SVRE}{M. Ehre, I. Papaioannou and D. Straub}

\title{Stein variational rare event simulation \thanks{Submitted to the editors DATE.
\funding{This project was supported by the German Research Foundation (DFG) through grants STR 1140/11-1 and PA2901/1-1.}}}

\author{Max Ehre\thanks{Engineering Risk Analysis Group, Technical University of Munich
  (\email{max.ehre@tum.de},\email{iason.papaioannou@tum.de},\email{straub@tum.de}).}
\and Iason Papaioannou\footnotemark[2] 
\and Daniel Straub\footnotemark[2].}

\usepackage{amsopn}
\DeclareMathOperator{\diag}{diag}

\usepackage[algo2e,ruled,lined]{algorithm2e}

\newcommand{\T}{\mathrm{T}}

\ifpdf
\hypersetup{
  pdftitle={\title{Estimating rare event probabilities with Stein variational gradient descent.}
},
  pdfauthor={M. Ehre, I. Papaioannou and D. Straub}
}
\fi

\Crefname{algocf}{Algorithm}{Algorithms}

\usepackage{titlesec}
\usepackage{tabto}

\titleformat{\paragraph}
{\normalfont\normalsize\bfseries}{\theparagraph}{1em}{}
\titlespacing*{\paragraph}
{0pt}{3.25ex plus 1ex minus .2ex}{1.5ex plus .2ex}

\begin{document}

\maketitle

\begin{abstract}
Accurately estimating rare event probabilities for systems subject to uncertainty and randomness is an inherently difficult task that calls for dedicated tools and methods. One way to improve estimation efficiency on difficult rare event estimation problems is to leverage gradients of the computational model representing the system under consideration to explore the rare event faster and more reliably. We present a novel approach for estimating rare event probabilities using such model gradients by drawing on a technique to generate samples from non-normalized posterior distributions in Bayesian inference -- the Stein variational gradient descent. We propagate samples generated from a tractable input distribution towards a near-optimal rare event importance sampling distribution by exploiting a similarity of the latter with Bayesian posterior distributions. Sample propagation takes the shape of passing samples through a sequence of invertible transforms such that their densities can be tracked and used to construct an unbiased importance sampling estimate of the rare event probability -- the Stein variational rare event estimator. We discuss settings and parametric choices of the algorithm and suggest a method for balancing convergence speed with stability by choosing the step width or base learning rate adaptively. We analyze the method's performance on several analytical test functions and two engineering examples in low to high stochastic dimensions ($d= 2 - 1500$) and find that it consistently outperforms other state-of-the-art gradient-based rare event simulation methods.
\end{abstract}

\begin{keywords}
	Rare event simulation,
	Stein variational gradient descent,
	Sequential importance sampling,
\end{keywords}

\begin{AMS}
  62L99, 62P30, 62J02, 65C05
\end{AMS}

\graphicspath{{figures/}{plots/}}
\section{Introduction and previous work}
\label{sec:intro}
The goal of rare event simulation is to simulate outcomes of a model that belong to a predefined event whose occurence probability is small ($\leq 10^{-3}$). 
Accurately estimating rare event and failure probabilities is relevant in a number of disciplines such as engineering design \cite{Faber2012}, structural reliability \cite{Lemaire2009}, chemical physics \cite{Allen2006}, finance \cite{Embrechts2013}, qeueing theory \cite{Asmussen2006}, traffic management or systems biology \cite{Rubino2009}.
In fields like engineering, finance or traffic management, the interest in such events is usually due to their associated severe adverse consequences, i.e., the events of interest are failure events \cite{Rubino2009,Ditlevsen1996,Lemaire2009}. 
\\~\\
Technically, computing rare event probabilities requires integrating over all states of the model input parameter distribution that trigger the event. 
Monte Carlo integration is inefficient for evaluating such tail integrals as, at a fixed estimation accuracy, the required number of samples scales inversely with the rare event probability.
Early approaches to efficiently approximate rare event probabilities were based on first- and second-order Taylor expansions of the hypersurface separating the rare event from the remaining input parameter space \cite{Rackwitz1978, Breitung1984,DerKiureghian2005}. The probability content of the resulting approximations can be computed with ease but these approaches are limited to relatively low parameter dimensions \cite{Valdebenito2010} and their accuracy can deteriorate quickly if the rare event boundary is strongly nonlinear.
Advanced sampling-based methods aim at reducing the number of required samples at constant estimation error compared to Monte Carlo. Proponents of such methods are classic variance reduction techniques like importance sampling \cite{Bucher1988,Engelund1993,Au1999} and its sequential variants \cite{Beaurepaire2013,Papaioannou2016}, in which importance sampling is applied repeatedly to sample from a sequence of distributions that converges towards an optimal (zero-variance) importance sampling density of  the rare event. Markov chain Monte Carlo (MCMC) schemes are used to propagate samples between subsequent distributions in the sequence \cite{Cerou2012,Papaioannou2016}. Alternatively, cross-entropy importance sampling \cite{Rubinstein1997, Rubinstein2017} and improved cross-entropy importance sampling (iCE) \cite{Papaioannou2019b} repeatedly solve a cross-entropy minimization problem over a parametrized distribution family to identify a similar sequence of intermediate distributions converging towards an optimal IS density that is subsequently used to sample the rare event. Recently, \cite{Uribe2020} introduced iCEred, which extends the applicability of the iCE approach to high-dimensional problem settings  by using gradient information of the rare event to identify a low-dimensional subspace in which the iCE algorithm is executed. 
These approaches are essentially sequential Monte Carlo (SMC) methods \cite{Doucet2001, DelMoral2006} that have been repurposed for exploring rare events rather than Bayesian posterior distributions. A popular variant of rare event-oriented SMC known as multilevel splitting \cite{Glasserman1999,Rubinstein2017} or subset simulation \cite{Au2001} is based on a particular (discontinuous) choice of the intermediate distributions.
The performance of subset simulation is largely determined by the efficiency and robustness of the chosen MCMC algorithm \cite{Papaioannou2015}.
If gradient information of the model is available, using gradient-based MCMC samplers such as Hamiltonian Monte Carlo (HMC) \cite{Duane1987,Neal2011} can improve the performance of subset simulation \cite{Wang2019b}. \cite{Papakonstantinou2023} proposed to use HMC for directly sampling from a near-optimal rare event importance sampling density.
\\~\\
Stein variational gradient descent (SVGD) is an algorithm for sampling from Bayesian posteriors without knowing the model evidence or -- more generally -- for sampling from non-normalized target distributions without knowing the normalization constant \cite{LiuWang2016}.
The principal idea is to pass a set of samples initially drawn from a tractable reference distribution through a sequence of invertible transforms that are chosen from a reproducing kernel Hilbert space such that the samples converge towards the target distribution. SVGD has grown to be a viable alternative to MCMC and has been successfully applied in various tasks connected to sampling from Bayesian posteriors such as uncertainty quantification in stochastic PDEs \cite{Zhu2018}, deep learning \cite{WangLiu2016} and training diversified mixture models \cite{Wang2019}. Theoretical analyses of SVGD are available in \cite{LiuQiang2017, Liu2019, Lu2019, Korba2020, Nuesken2023}. Several relevant improvements have been made to SVGD such as using surrogate models of the target to drive a gradient-free version of SVGD \cite{Han2018} and subspace SVGD methods \cite{Chen2020,Liu2022} that aim at resolving a variance degeneracy issue faced when using SVGD in high parameter dimensions \cite{Ba2022}.
\\~\\
In this contribution, we introduce the Stein variational rare event (SVRE) estimator. The SVRE estimator is an importance sampling estimator that relies on repurposing SVGD for sampling near-optimal, smoothened rare event importance densities. Following \cite{Han2017}, we obtain an unbiased rare event probability estimate using the final generation of SVRE-generated samples. 
We also discuss an approximation that allows choosing the base learning rate adaptively such as to strike a good balance between accuracy and convergence speed. 
\\~\\
The manuscript is organized as follows: \cref{sec:background}  introduces the rare event estimation problem and briefly covers importance sampling estimates of rare event probabilities in \cref{sec:rare_events}. In the remainder of \cref{sec:background}, we give a review of the principles underlying SVGD and the original SVGD algorithm. \cref{sec:rare_event_svgd} lays out our method for estimating rare event probabilties with the SVRE estimator.  \cref{sec:numerical} provides empirical results on a number of test functions and engineering examples and conclusions are given in  \cref{sec:conclusion}. 
\section{Background}
\label{sec:background}
\subsection{Rare event estimation}
\label{sec:rare_events}
Let $\bm{X}$ be a $d$-dimensional random vector that is absolutely continuous with respect to the Lebesgue measure and has joint probability density function (PDF) $p_0(\bm{x})$. Let $\mathcal{X}$ denote the outcome space $\bm{X}$ maps to so that each realization of the random vector $\bm{x} \in \mathcal{X}$ and let $p_0(\bm{x}) > 0,~ \bm{x} \in \mathcal{X}$.
We define a rare event as $\mathcal{F} = \{\bm{x}: g(\bm{x}) \le 0\}$, where $g(\bm{x})$ is a performance function, which is $\leq 0$ if $\bm{x}$ belongs to the rare event. 
In structural reliability analysis, the rare event corresponds to a failure event and the performance function is also known as limit state function (LSF). The probability of the failure event is
\begin{equation}
\label{e:pf}
p_{\mathcal{F}} = \int_{\mathcal{F}}  p(\bm{x})\mathrm{d}\bm{x} = \int_{\mathcal{X}} \mathrm{I}[g(\bm{x}) \le 0] p_0(\bm{x})\mathrm{d}\bm{x} = \mathbb{E}_{p_0}\left[\mathrm{I}(g(\bm{X}) \le 0)\right] ,
\end{equation}
where the indicator function $\mathrm{I}[\cdot]$ equals 1 if the condition in the argument is true and 0 otherwise.
The  Monte Carlo estimator of \cref{e:pf} using $n$ independent samples from $p_0(\bm{x})$ has coefficient of variation $\sqrt{(1-p_{\mathcal{F}})/(n p_{\mathcal{F}})}$, which becomes large for small $p_{\mathcal{F}}$, i.e., for rare event probabilities.
Importance sampling (IS) belongs to a class of methods known as variance reduction techniques and can improve the situation significantly by sampling from a density $h$ rather than $p_0$, where $h$ should cover the failure event better than $p_0$. 
Let $h$ be a PDF such that $h\left(\bm{x}\right) > 0$ whenever $g\left(\bm{x}\right) \le 0$. Then, one can rewrite \cref{e:pf} as
\begin{equation}
	\label{e:pf_is}
	p_{\mathcal{F}} =\int_{\mathcal{X}} \mathrm{I}[g\left(\bm{x}\right) \le 0] \overbrace{\frac{p_0(\bm{x})}{h(\bm{x})}}^{ \vcentcolon = w (\bm{x})} h(\bm{x}) \mathrm{d}\bm{x}
	= \mathbb{E}_{h}\left[\mathrm{I}(g(\bm{X}) \le 0) \omega(\bm{X})\right],
\end{equation}
which leads to the (unbiased) IS estimator
\begin{equation}
	\label{eq:phat_is}
	\widehat{p}_{\mathcal{F}} = \frac{1}{n} \sum\limits_{k=1}^{n} \mathrm{I}[g(\bm{x}_k) \le 0] w(\bm{x}_k),~~~\bm{x}_k \sim h.
\end{equation}
The efficiency of IS depends on the choice of $h$, the selection of which is nontrivial and an ongoing research topic \cite{Papaioannou2016,WangSong2016,Chiron2023, Uribe2020}. An optimal importance density $h^*$, whose associated estimator's coefficient of variation is zero, reads
\begin{equation}
	\label{eq:hopt}
	h^*(\bm{x}) = \frac{\mathrm{I}[g(\bm{x}) \le 0] p_0(\bm{x})}{p_{\mathcal{F}}}.
\end{equation}
The normalizing constant of $h^*$ is $p_{\mathcal{F}}$  -- the sought rare event probability. Hence, sampling from $h^*$ requires sampling from a PDF without knowing its normalizing constant. 
\subsection{Variational inference with reversible transformations}
\label{sec:vi}
Variational inference (VI) is a class of methods that approximately solve Bayesian inference through solving an equivalent optimization problem. The basic idea is to approximate the non-normalized posterior $\tilde{p} = p(x) Z$ with $Z = \int \tilde{p}(\bm{x}) \mathrm{d}\bm{x}$ using another distribution $q \in \mathcal{Q}$ \cite{Blei2017}. $\mathcal{Q}$ is often chosen as a parametrized family of distributions. The goal is to find a $q$ that is most similar to $p$ in some sense. Usually, the Kullback-Leibler-divergence is used as a discrepancy measure such that the VI problem reads
\begin{equation}
	q^*(\bm{x}) = \argmin_{q \in \mathcal{Q}} D_{\mathrm{KL}}(q || p) = \argmin_{q \in \mathcal{Q}}  \mathbb{E}_q[ \log q] - \mathbb{E}_q[ \log \tilde{p}] + \log Z.
\end{equation}
$\log Z$ is independent of $q$ and hence can be neglected for solving the optimization problem.
A key challenge in any VI method is the choice of $\mathcal{Q}$, which should be simple enough to keep the optimization problem tractable but sufficiently expressive to allow for approximating complex posterior distributions.
Parametric disitribution families often fail to accomplish the latter. SVGD can be viewed as a VI method, in which $\mathcal{Q}$ is defined implicitly as the set of distributions obtained by pushing a random vector $\bm{X} \sim q_0$ through a sequence of $T$ bijective transforms $\{\mathcal{T}_t: \mathcal{X} \to \mathcal{X} \}_{t=1}^T$ \cite{LiuWang2016}. For any bijective transform $\mathcal{T}$, we call the distribution of  $ \bm{Z} = \mathcal{T}( \bm{X})$ with $\bm{X} \sim q_0$ the pushforward of $q_0$ under $ \mathcal{T}$ and denote it as $ \mathcal{T}_{\#}q_0$. Further, we denote the pushforward of $q_0$ under $\mathcal{T}_t \circ \dots \circ  \mathcal{T}_1$ as $q_t \vcentcolon = (\mathcal{T}_t\circ \dots \circ  \mathcal{T}_1)_{\#} q_0$. 
Within SVGD, the posterior distribution is hence approximated in $\mathcal{Q} = \{\{\mathcal{T}_t\}_{t=1}^T : (\mathcal{T}_T\circ \dots \circ  \mathcal{T}_1)_{\#} q_0\}$, where the $\{\mathcal{T}_t\}_{t=1}^T$ are chosen as
\begin{equation} 
\mathcal{T}_t(\bm{x}) = \bm{x} + \epsilon \phi_t(\bm{x})~(t = 1,\dots,T).
\end{equation}
$\phi_t: \mathcal{X} \to \mathcal{X}$ is a smooth, bounded vector-valued function and $\epsilon$ is a learning rate. $\phi_t$ can be interpreted as a velocity field along which particles living in $\mathcal{X}$ flow for a time $\epsilon$. Choosing  $\epsilon < 1$ ensures that $\mathcal{T}_t(\bm{x}) $ is close to the identity map and bijective such that by the transformation of variables theorem
\begin{equation} 
	\label{eq:q_T}
	q_T(\bm{x}^T) = q_0(\bm{x}^0) \prod_{t=0}^{T-1} |\mathrm{det} \nabla \mathcal{T}_t(\bm{x}^{t})|^{-1},
\end{equation}
where $\bm{x}^{t} = \mathcal{T}_t \circ \dots \circ  \mathcal{T}_1 (\bm{x}_0)$ and $\bm{x}_0 \sim q_0$.
Starting from $q_0$, the main task in SVGD is to find a sequence of velocity fields that minimize $D_{\mathrm{KL}}(q_t || p)$ in each step $t=1,\dots,T$. This corresponds to solving the following optimization problem in each step \cite{LiuWang2016}:
\begin{align}
	\label{eq:minimization}
	\begin{split}
	\phi_t(\bm{x}) &= \argmin_{\phi \in \mathcal{P}} \nabla_{\epsilon}  D_{\mathrm{KL}}(q_t || p)\big|_{\epsilon=0} \\
	&= \argmin_{\phi \in \mathcal{P}} \nabla_{\epsilon}  D_{\mathrm{KL}}(q_{t-1} || \mathcal{T}_{t~\#}^{-1} p)\big|_{\epsilon=0} \\
	&=  - \argmin_{\phi \in \mathcal{P}} \mathbb{E}_{q_{t-1}}[ \nabla_{\epsilon} \log \mathcal{T}_{t~\#}^{-1} p)]\big|_{\epsilon=0} \\
	&=  \argmax_{\phi \in \mathcal{P}} \mathbb{E}_{q_{t-1}}[ \nabla_{\bm{x}} \log p( \mathcal{T}_t(\bm{x}))^\T \phi_t(\bm{x}) + \text{trace}((\mathbf{I} + \epsilon \nabla_{\bm{x}} \phi_t(\bm{x}))^{-1} \cdot  \nabla_{\bm{x}} \phi_t(\bm{x}) )]\big|_{\epsilon=0} \\
	&=  \argmax_{\phi \in \mathcal{P}} \mathbb{E}_{q_{t-1}}[\text{trace}(\mathcal{A}_p \phi_t(\bm{x}))],
	\end{split}
\end{align}
where $\mathcal{A}_p \phi_t(\bm{x}) = \nabla_{\bm{x}} \log p(\bm{x})^\T \phi_t(\bm{x}) + \nabla_{\bm{x}}  \phi_t (\bm{x}) $ is a so-called Stein operator and $\mathcal{P}$ is some vector-valued normed function space containing all possible choices for $\phi$. Choosing $\mathcal{P}$ as a \textit{reproducing kernel Hilbert space} (RKHS) will facilitate solving \cref{eq:minimization} and thus we introduce kernels and the concept underlying RKHS next.  
\subsection{Kernels and RKHS}
Let $k: \mathcal{X} \times \mathcal{X} \to \mathbb{R}$ be a symmetric, positive definite function. The closure of  $\{f: f(\bm{x}) = \sum_{i=1}^\infty a_i k(\bm{x},\bm{y}_i)\}$ is the unique reproducing kernel Hilbert space of $k$, $\mathcal{H}$ \cite{Aronszajn50}.  Since $k$ is positive definite, $\mathcal{H}$ can be equipped with the inner product $\langle f,g \rangle_{\mathcal{H}} = \sum_{i=1}^\infty \sum_{j=1}^\infty a_i b_j k(\bm{x}_j,\bm{y}_i)$ with $g(\bm{y}) = \sum_{j=1}^\infty b_j k(\bm{x}_j,\bm{y}) \in \mathcal{H}$ and norm $\lVert f \rVert_{\mathcal{H}} = \sqrt{ \langle f,f \rangle_{\mathcal{H}}}$. Further, let \smash{$\mathcal{H}^d = \bigtimes_{i=1}^d \mathcal{H}$} be a product Hilbert space of $d \times 1$-vector-valued functions $\bm{f} = [f_1, \dots, f_d]^\T$ with $\{f_i  \in \mathcal{H}\}_{i=1}^d$ and inner product \smash{$\langle \bm{f}, \bm{g} \rangle_{\mathcal{H}^d} = \sum_{i=1}^d \langle f_i,g_i \rangle_{\mathcal{H}}$}, where  $\bm{g} = [g_1, \dots, g_d ]^\T \in \mathcal{H}^d$. The norm in $\mathcal{H}^d$ is defined as \smash{$\lVert \bm{f} \rVert_{\mathcal{H}^d} =  (\sum_{i=1}^d \langle f_i,f_i \rangle_{\mathcal{H}})^{1/2} =  (\sum_{i=1}^d \lVert f_i \rVert_{\mathcal{H}}^2)^{1/2}$}. $k$ can be always expressed in terms of a (non-unique) map $\phi: \mathcal{X} \to \mathcal{H}^d$ (known as the feature map) as $k(\bm{x},\bm{y}) = \langle \phi(\bm{x}),  \phi(\bm{y}) \rangle_{\mathcal{H}^d}$. 
The two key properties of the RKHS $\mathcal{H}^d$ are \cite{Steinwart2008}:\\
\begin{enumerate}
	\item $k(\bm{x}, \cdot) \in \mathcal{H}^d$.
	\item $f(x) = \langle f(\cdot), k(\bm{x}, \cdot)\rangle_{\mathcal{H}^d }$.
\end{enumerate}
~\\
The first property states that the canonical map $k(\bm{x},\cdot)$ is in the RKHS and is implied in the definition of the inner product of the RKHS. The second property is the reproducing property and states that the evaluation operator of any function in the RKHS at $\bm{x}$ can be expressed as an inner product of said function with the canonical map evaluated at $\bm{x}$. Effectively, this is a smoothness constraint on the elements of the RKHS and the key to obtain a closed-form solution of \cref{eq:minimization} once the search space is chosen as $\mathcal{H}$.
\subsection{Stein discrepancy and optimal KL descent}
Let $q$ and $p$ be PDFs of distributions supported on $\mathcal{X}$ and assume $\phi(\bm{x})$ satisfies
\begin{equation*}
	\int_{\mathcal{X}} \nabla_{\bm{x}} ( p(\bm{x}) \phi^\T(\bm{x}) ) \mathrm{d}\bm{x} = \bm{0}.
\end{equation*}
Then, $\phi$ is said to be in the Stein class of $p$ and
\begin{equation}
\label{eq:SteinsIdentitiy}
	\mathbb{E}_q [\mathcal{A}_p \phi(\bm{x})] = \bm{0} \text{ iff } q = p \text{ with } \mathcal{A}_p \phi_t(\bm{x}) = \nabla_{\bm{x}} \log p(\bm{x})^\T \phi_t(\bm{x}) + \nabla_{\bm{x}}  \phi_t (\bm{x}) 
\end{equation}
as defined below \cref{eq:minimization}.
For $q = p$,  \cref{eq:SteinsIdentitiy} is known as Stein's identity and maximizing the violation of Stein's identity over $\mathcal{P}$ if $q \neq p$ gives rise to a statistical divergence measure known as Stein discrepancy $\mathbb{D}(p,q)$, which expresses how different $q$ and $p$ are \cite{Gorham2015}:
\begin{equation}
	\label{eq:SteinsDiscrepancy}
	\mathbb{D}(p,q) \vcentcolon = \max\limits_{\phi \in \mathcal{P}}  	\mathbb{E}_q [\text{trace} (\mathcal{A}_p \phi(\bm{x}))].
\end{equation}
The choice of $\mathcal{P}$ -- known as the Stein set -- plays a similar role as the choice of $\mathcal{Q}$ in VI: very general choices of $\mathcal{P}$ (e.g., all Lipschitz-differentiable functions) render the above maximization intractable whereas too specific choices of $\mathcal{P}$ may overly constrain the search space and lead to $\mathbb{D}(p,q) = 0$ even if $q \neq p$ \cite{LiuLee2016}. 
\\~\\
Constraining $\mathcal{P}$ to a unit ball in $\mathcal{H}^d$ and assuming $k$ is in the Stein class of $p$, one may use the reproducing property -- namely $f(\bm{x}) = \langle f(\cdot), k(\bm{x},\cdot) \rangle_{\mathcal{H}^d}$ and $ \nabla_{\bm{x}} f(\bm{x}) = \langle f(\cdot),  \nabla_{\bm{x}}  k(\bm{x},\cdot) \rangle_{\mathcal{H}^d}$ -- to obtain a \emph{kernelized Stein discrepancy} \cite{Chwialkowski2016,LiuLee2016}
\begin{equation}
	\label{eq:SteinsDiscrepancyHilbert}
	\mathbb{S}(p,q) = \max\limits_{\substack{\phi \in \mathcal{H}^d \\ \lVert \phi \rVert_{\mathcal{H}^d} \leq 1}}\mathbb{E}_{q}[\mathrm{trace}(\mathcal{A}_p \phi(\bm{x}))] =  \max_{\substack{\phi \in \mathcal{H}^d \\ \lVert \phi \rVert_{\mathcal{H}^d} \leq 1}} \langle \phi(\cdot), \mathcal{A}_p k(\bm{x},\cdot) \rangle_{\mathcal{H}^d}.
\end{equation}
%
Following from the right-hand side in \cref{eq:SteinsDiscrepancyHilbert}, we can compute $	\mathbb{S}(p,q)$ by choosing $\phi \propto \mathcal{A}_p k(\bm{x},\cdot)$ and normalizing such that
\begin{equation}
	\label{eq:opt_map}
	\mathbb{S}(p,q_{t-1}) = \mathbb{E}_{q_{t-1}} \left[\mathrm{trace}\left(\mathcal{A}_p \frac{\phi^\star_t(\bm{x})}{\lVert \phi^\star_t \rVert_{\mathcal{H}^d}}\right)\right]  \text{ with } \phi^\star_t(\cdot) = \mathbb{E}_{q_{t-1}} [\nabla_{\bm{x}} \log p(\bm{x})^\T k(\bm{x},\cdot) + \nabla_{\bm{x}}. k(\bm{x},\cdot) ]
\end{equation}
This implies that 
\begin{equation}
	\mathbb{S}(p,q_{t-1}) = \sqrt{\mathbb{E}_{q_{t-1}} [\mathrm{trace}(\mathcal{A}_p^{\bm{x}} \mathcal{A}_p^{\bm{y}} k(\bm{x},\bm{y}))]} =  \langle \phi^\star_t, \mathcal{A}_p k(\bm{x},\cdot) \rangle_{\mathcal{H}^d} =  \lVert \phi^\star_t \rVert_{\mathcal{H}^d},
\end{equation}
where $\mathcal{A}_p^{\bm{x}}$ acts on the first argument of $k$ and $\mathcal{A}_p^{\bm{y}}$ acts on the second argument of $k$.
Comparing \cref{eq:SteinsDiscrepancy} and 	\cref{eq:minimization} reveals that transporting $\bm{x} \sim q_{t-1}$ along $\phi^\star_t$ for a step width $\epsilon_t$ reduces the KL divergence with the target $p$ by an amount $\mathbb{S}(p,q_{t-1})$. 
\\~\\
This facilitates an iterative procedure that succesively drives the sequence of variational distributions $\{q_t\}_{t=1}^T$ closer to $p$. The expectation $\mathbb{E}_{q_{t-1}}[\cdot]$ in \cref{eq:opt_map} is approximated with a finite number of samples/particles that are initially drawn from the first (tractable) variational PDF $q_0$ and subsequently propagated. $q_0$ is initialized with $p_0$. The original SVGD algorithm is outlined in \cref{alg_1}.
\vspace{.5cm}~\\
	\begin{algorithm2e}[H]
		\caption{SVGD \cite{LiuWang2016}}
		\label{alg_1}
		\SetKwInOut{Input}{Input}
		\SetKwInOut{Output}{Output}
		
		\Input{target PDF $p(\bm{x})$, kernel $k(\bm{x},\bm{x}')$, $n$ prior samples $\{\bm{x}_i^0 \sim p_0\}_{i=1}^n$, \# of steps $T$, step size $\epsilon_t$}
		\Output{$n$ posterior samples $\{\bm{x}_i^T \sim p\}$}
		\For{$t = 0,\dots,T$}{
			$\bm{x}_i^{t+1} = \bm{x}_i^t + \epsilon_t \widehat{\phi}_t^*(\bm{x}_i^t)$   with   
			$ \widehat{\phi}_t^*(\bm{y}) = \frac{1}{n} \sum\limits_{i=1}^n k(\bm{x}_i^t,\bm{y}) \nabla_{\bm{x}} \log p(\bm{x}_i^t) + \nabla_{\bm{x}} k(\bm{x}_i^t,\bm{y})$
		}
	\end{algorithm2e}
\vspace{.5cm}
\section{SVRE: Stein variational rare event estimation}
\label{sec:rare_event_svgd}
\subsection{Method}
\label{sec:method}
\begin{figure}
	\centering
	\includegraphics[width=0.75\textwidth]{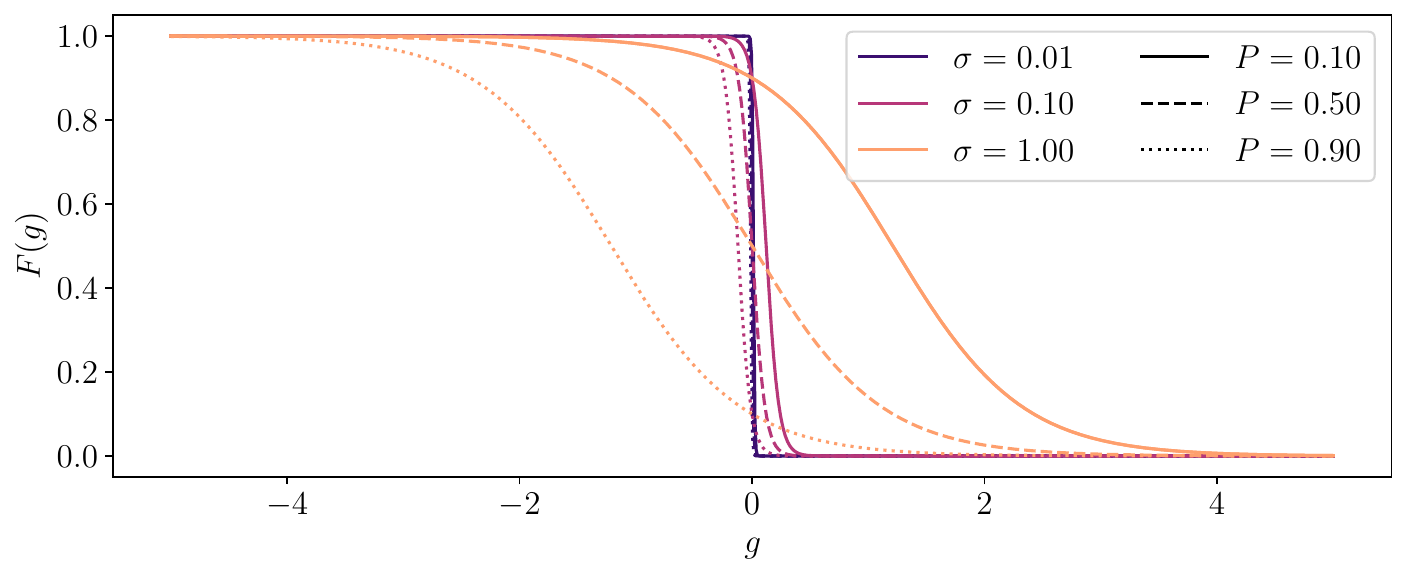}
	\caption{Smoothened indicator function for different choices of the smoothing parameters $P, \sigma$.}
	\label{fig:smooth_indicator}
\end{figure}

In this section we formulate our method that aims at repurposing the original SVGD to simulate samples from a given rare event $\mathcal{F}$ and subsequently estimating the associated rare event probability with importance sampling based on the final set of samples. To this end we exploit a similarity between a generic Bayesian posterior distribution
\begin{equation}
	\label{eq:bayes}
	p_{\mathrm{post}}(\bm{x}) = \frac{L(\bm{x}) p_{\mathrm{prior}}(\bm{x})}{Z}
\end{equation}
and the optimal rare event importance sampling density in \cref{eq:hopt}. In particular, the posterior $p_{\mathrm{post}}(\bm{x})$ and the optimal rare IS density $h^*(\bm{x})$ are both products of two pointwise known functions and an unknown normalization constant. Therein, the likelihood $L(\bm{x})$ corresponds to the indicator $\mathrm{I}[g(\bm{x}) \leq 0]$, the prior $p_{\mathrm{prior}}(\bm{x})$ corresponds to the nominal PDF $p_0(\bm{x})$ and the evidence $Z$ corresponds to the rare event probability $p_{\mathcal{F}}$. This analogy has been leveraged to repurpose Bayesian inference approaches for sampling from rare event optimal IS densities in \cite{Papaioannou2016,Uribe2020, Wagner2022, Ehre2023}.
\\~\\
In order to avoid vanishing gradients and regularize the inference problem, it is necessary to smoothen the rare event indicator function $\mathrm{I}[g(\bm{x}) \leq 0]$. Here, we use the logistic CDF to obtain a smoothened indicator of the form
\begin{equation}
	\label{eq:smooth_indicator}
	F(\bm{x}) = \frac{1}{2}\left(1 + \tanh \left(- \frac{\pi}{\sqrt{3}} \frac{ \mu + g(\bm{x}) }{2 \sigma}\right)\right)~~~\text{with}~~~\nabla_{\bm{x}} \log F(\bm{x}) = - \frac{\pi \nabla_{\bm{x}} g(\bm{x})}{2\sqrt{3}\sigma} \left(1 - \tanh\left( - \frac{\pi}{\sqrt{3}}  \frac{ \mu + g(\bm{x}) }{2 \sigma} \right)\right).
\end{equation}
$\mu$ is the mean of the logistic RV and controls the location of the smooth indicator. The choice $\mu = - \sqrt{3} \sigma / \pi \log(P/(1-P))$ places probability mass $P$ in the failure domain. This approach is also used in \cite{Papakonstantinou2023}. We adopt their setting and select $P = 0.9$.
$\sigma$ is the standard deviation of the logistic RV and acts as a smoothing parameter. 
\cref{fig:smooth_indicator} depicts the smoothened indicator for different choices of $\mu$ and $\sigma$ and shows that the choice of $\mu$ has negligble influence when $\sigma \leq 0.01$ (we chose $\sigma = 0.001$ for all our tests).
Plugging \cref{eq:smooth_indicator} for the indicator in \cref{eq:hopt}, we obtain a smooth rare event IS density that can be defined as a target PDF:
\begin{equation}
	\label{eq:hopt_svgd}
	p(\bm{x}) \vcentcolon= \frac{F(\bm{x})  p_0(\bm{x})}{\tilde{p}_{\mathcal{F}}}~~~\text{with}~~~\tilde{p}_{\mathcal{F}} = \int_{\mathcal{X}} F(\bm{x})  p_0(\bm{x})  \mathrm{d}\bm{x}.
\end{equation}
Consequently, the optimal map in step $t$ reads
\begin{equation}
	\label{eq:opt_smooth_indicator_map}
	 \phi^\star_t(\cdot) = \mathbb{E}_{q_{t-1}} [(\nabla_{\bm{x}} \log F(\bm{x}) + \nabla_{\bm{x}} \log p_0(\bm{x}))^\T k(\bm{x},\cdot) + \nabla_{\bm{x}}   k(\bm{x},\cdot) ].
\end{equation}
Upon running SVGD for $T$ iterations in order to simulate $n$ samples from $p(\bm{x})$, the obtained samples $\{\bm{x}_i^{T}\}_{i=1}^n$ follow the final variational IS distribution $q_T$. The importance sampling estimate of the rare event probability reads
\begin{equation}
	\label{eq:phat_svgd}
	\widehat{p}_{\mathcal{F}}^{\mathrm{SVRE}}= \frac{1}{n} \sum\limits_{i=1}^{n} \mathrm{I}[g(\bm{x}^T_i) \le 0] \frac{p_0(\bm{x}^T_i)}{q_T(\bm{x}^T_i)}~~~\text{with}~~~\bm{x}^T_i \sim q_T,
\end{equation}
where $q_T$ is evaluated by tracking the Jacobians of all the $\{\mathcal{T}_t\}_{t=1}^T$ using \cref{eq:q_T}. 
Assuming i.i.d. samples, \cref{eq:phat_svgd} is unbiased and an estimate of the coefficient of variation of $\widehat{p}_{\mathcal{F}}^{\mathrm{SVRE}}$ is given as \cite{Owen2013}
\begin{equation}
	\label{eq:deltahat_svgd}
	\textstyle
	\widehat{\delta}_{\mathcal{F}}^{\mathrm{SVRE}}
	= \sqrt{ \frac{\sum_{i=1}^{n} w_i^2}{\left(\sum_{i=1}^{n} w_i\right)^2} - \frac{1}{n} }  ~~~\text{where}~~~w_i \vcentcolon= \mathrm{I}[g(\bm{x}^T_i) \le 0] \frac{p_0(\bm{x}^T_i)}{q_T(\bm{x}^T_i)},
\end{equation}
\subsection{Algorithm and implementation details}
\label{sec:algorithm}
\subsubsection*{Sample split}
\label{sec:sample_split}
\begin{figure}[h!]
	\centering
	\subfloat[][]{\includegraphics[width=0.5\textwidth]{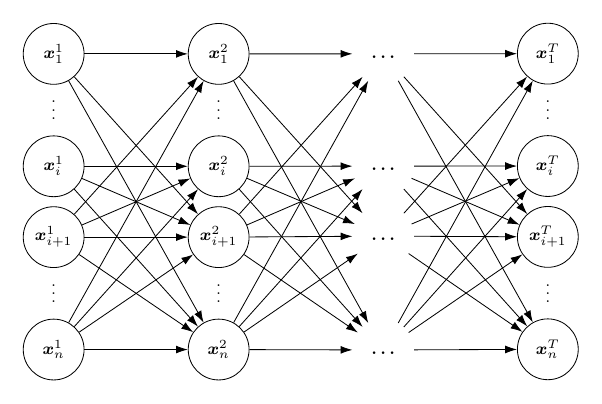}\label{fig:particle_system_0}}
	\subfloat[][]{\includegraphics[width=0.5\textwidth]{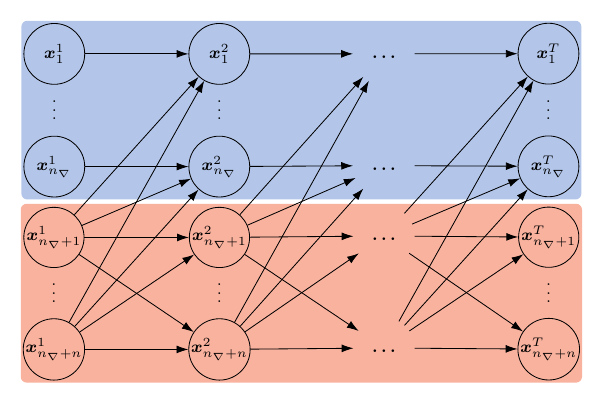}\label{fig:particle_system_1}}
	\caption{SVGD particle systems without (a) and with (b) separating inducing and estimation samples (inspired by \cite{Han2017}). In case b), the first $n_{\nabla}$ samples are used for constructing $\mathcal{T}_t$ (blue box). The remaining $n$ particles are propagated along  $\mathcal{T}_t$ and used for estimating $p_{\mathcal{F}}$.}
	\label{fig:particle_systems}
\end{figure}

Conceptually, the rare event setting presented in this work is somewhat different from the Bayesian inference applications tackled in the machine learning context, in which SVGD was conceived originally. In particular, evaluating the gradient of the log-likelihood or -- in the context of rare events -- the log-smooth indicator function, requires evaluating LSF gradients, which dominates computational cost in an engineering context. This is due to the fact that the computational model behind $g$ is usually computationally expensive, e.g., a finite element or finite volume approximation of the solution to a set of partial differential equations. 
In the original SVGD, all $n$ samples are used for two tasks simultaneously:
~\\
\begin{enumerate}
	\item estimating the expectation in $\phi_t$ in each step (see \cref{eq:opt_map})
	\item tracking the sequence of variational IS densities $\{q_t\}_{t=1}^T$ (by propagation through $\{\mathcal{T}_t\}_{t=1}^T$)
 \end{enumerate}	
~\\
It is, however, straightforward to evaluate the empirical expectation using only a subset of all available samples, which has first been proposed in \cite{Han2017}. This is useful as LSF gradients are required only for task 1. Once the subset of samples used for estimating the expectation is fixed, all samples are propagated through the $T_t$ (task 2), which only requires kernel evaluations and neither LSF values nor LSF gradients. This bears a signficant potential for reducing the overall number of required LSF gradient evaluations. In the context of importance sampling, performing task 1 on a subset of all samples has an additional advantage: the subset of all samples not used in task 1 remain independent and hence using only these samples in \cref{eq:phat_svgd} renders the IS estimate unbiased and facilitates using \cref{eq:deltahat_svgd} for computing the estimator coefficent of variation.
Assuming a total $n + n_\nabla$ samples, we call the $n_{\nabla}$ samples used for task 1 \emph{inducing samples}, and the remaining $n$ samples \emph{estimation samples}.
\cref{fig:particle_systems} shows a schematic illustrating this modification.
\subsubsection*{Normalization schemes}
\cite{LiuWang2016} suggest replacing the normalization of the optimal map in \cref{eq:opt_map} with an adagrad/RMSProp-like normalization term that accumulates gradient histories in an exponential moving average.
That is, the learning rate in each step $\epsilon_t$, becomes a $1 \times d$ vector of individual learning rates along each of the $d$ dimensions and depends on $\bm{x}$, i.e., is different for each particle. The transformation at step $t$ then reads
\begin{equation} 
	\mathcal{T}_t(\bm{x}) = \bm{x} + \epsilon_t(\bm{x}) \odot \phi_t(\bm{x})~(t = 1,\dots,T).
\end{equation}
A more detailed description of RMSProp and the associated shape of $\epsilon_t(\bm{x})$ is detailed in \cref{appendix:B}.
 In the context of SVRE, we find this choice is effective for sampling from multimodal target IS densities, i.e., when tackling problems with multiple relevant failure domains. For unimodal targets, normalizing the optimal map with respect to $\ell_2$-norm in every step without any regard for gradient histories can outperform RMSProp. In particular, for problems with a dominant variable and nearly axis-parallel streamlines (e.g., the plate example in \cref{sec:plate}), RMSProp tends to inflate the influence of less relevant variables and push samples in the wrong direction. In \cref{appendix:B}, we detail both normalization schemes and how they affect the Jacobian $\nabla_{\bm{x}} \mathcal{T}_t(\bm{x})$. Unlike in \cite{Han2017}, 
 we account for the influence of $\epsilon_t(\bm{x})$ on the Jacobian of the optimal map and derive the exact Jacobians under non-constant learning rates.
\subsubsection*{Learning rate}
A base learning rate $\bar{\epsilon}_t$ is used to control the global step width independent of the normalization scheme, i.e., $\epsilon_t(\bm{x}) = \bar{\epsilon}_t / f(\bm{x})$, where $f(\cdot)$ represents the normalization scheme (see \cref{appendix:B}).
In the original SVGD algorithm, the learning rate $\bar{\epsilon}_t$ is chosen conservatively small, e.g., $ \bar{\epsilon}_t  = 0.001$, which frequently leads to SVGD running for several hundred iterations until convergence. This in turn increases the overall number of required log-likelihood gradient evaluations. As discussed, in the rare event setting with an underlying expensive-to-evaluate computational model, model gradient evaluations dominate computational cost.
Hence, our goal is to make $ \bar{\epsilon}_t $ as large as possible such that the total number of required LSF gradient evaluations becomes minimal, all while not inflating $ \bar{\epsilon}_t $ to the extent of compromising the invertibility of $\mathcal{T}_t$. In general, an $\bar{\epsilon}_t$ striking a good balance between both conflicting goals will depend on the shape of $g$. Depending on the normalization scheme, our studies show values $ \bar{\epsilon}_t  \in [0.25,1.0]$ to be appropriate choices that lead to convergence in $\mathcal{O}(- \log p_{\mathcal{F}})$ steps.
\\~\\
Alternatively, one can linearize the determinant of the Jacobian $\nabla_{\bm{x}} \mathcal{T}_t(\bm{x})$ in $\bar{\epsilon}_t$ and maximize $\bar{\epsilon}_t$ in a corridor such that all Jacobian determinants (at all samples) are bounded away from zero and below an upper threshold. This enables maximizing convergence speed while guaranteeing computational stability. Moreover, it reduces the computational effort associated with computing the determinant from $\mathcal{O}(n^3)$ to $\mathcal{O}(n)$. Details are provided in \cref{appendix:trace_approx}.
\subsubsection*{Choice of kernel/bandwidth selection}
In \cite{LiuWang2016}, the authors suggest using an isotropic Gaussian kernel 
\begin{equation}
		k(\bm{x},\bm{y}) = \exp\left\{ - \frac{\lVert \bm{x} - \bm{y} \rVert^2}{2 \ell ^2}\right\}
\end{equation}
with a single bandwidth parameter $\ell$. They suggest selecting $\ell$ such that at any sample $\bm{x} _i$, the weights of all neigboring log-likelihood gradients balance with its own log-likelihood gradient, i.e., $\sum_{j \neq i} k(\bm{x}_i,\bm{y}_j) =  k(\bm{x}_i,\bm{y}_i) = 1$. For an explicit choice of $h$, they approximate $\sum_{i \neq j} k(\bm{x}_i,\bm{y}_j) \approx n \exp ( - \text{median}(\lVert \bm{x} - \bm{y} \rVert)^2 / (2\ell^2))$ which yields $\ell^2 = \text{median}(\lVert \bm{x} - \bm{y} \rVert)^2 / (2 \log n)$. Our experiment suggests that -- in combination with RMSProp normalization -- this is an advantageous choice for multimodal targets. For unimodal targets -- for which we use $\ell_2$ normalization -- setting $\ell$ to a generic large value performs well (e.g.,  $\ell=10$ in standard normal space). The implication of this choice is that the effective gradient field will be an average of all $\log p(\bm{x}_i)$ - values shared by all samples. Hence all samples move identically such that the initial samples are shifted towards the rare event region like a rigid body. Therefore, if $p_0$ belongs to the exponential family, SVRE behaves like an exponential tilting IS estimate whose tilting parameter is the endpoint of the samples' mean trajectory.
\\~\\
Alternative choices of the kernel are possible but have not been explored in this work. In particular, kernels with slowly decaying tails such as the inverse multiquadratic kernel have been found to offer theoretical advantages over kernels with exponentially decaying tails in Stein discrepancy-driven sampling methods \cite{Gorham2017}. 
\subsubsection*{Stopping criterion}
In \cite{LiuWang2016}, no stopping criterion is used for the original SVGD. Likewise, for many SVGD descendants, stopping criteria and convergence checks are not discussed. Instead, these algorithms are executed for a predefined number (typically several hundred) of iterations. In order to minimze the number of iterations required, we terminate our algorithm once the empirical coefficient of variation of the importance sampling weights $\widehat{\delta}_w = \sqrt{n} \widehat{\delta}_{\mathcal{F}}^{\mathrm{SVRE}}$ drops below a prescribed threshold $\delta_{\mathrm{thresh}}$. We choose $\widehat{\delta}_w$ as it measures how close the sampling distribution and the target distribution are to one another independent of sample size.
As such it is closely connected to the relative effective sample size $\mathrm{rESS} = 1 / (1 + \widehat{\delta}_w^2)$ \cite{Latz2018}, which is another well-established performance measure for importance sampling distributions. 
\cite{Au2003} show that bounding the coefficient of variation of the IS failure estimate $\widehat{p}_{\mathcal{F}}$ from \cref{eq:phat_is} -- and therefore bounding the coefficient of variation of the weights $w$ -- is related to bounding the Kullback-Leibler-divergence from the sampling distribution $h$ to the target $p$. Applied to the SVRE, the relationship reads
\begin{equation}
	\sqrt{\exp\{D_{\mathrm{KL}}(p || q_t) \} - 1} \leq \widehat{\delta}_{\mathcal{F}}^{\mathrm{SVRE}}.
\end{equation}
\subsubsection*{Algorithm}~\\
A pseudocode of the SVRE method and estimator is provided in \cref{alg_2}.
	\begin{algorithm2e}[H]
	\caption{Stein variational rare event estimator (SVRE)}
	\label{alg_2}
	\SetKwInOut{Input}{Input}
	\SetKwInOut{Output}{Output}
	
	\Input{LSF $g(\bm{x})$, smoother $F(\bm{x})$ with $\mu$, $\sigma$, kernel $k(\cdot,\cdot)$, input PDF $p_0$, $\delta_{\mathrm{thresh}}$, $n$, $n_\nabla$}
	\Output{failure estimate $\widehat{p}_\mathcal{F}$, estimator c.o.v. $\widehat{\delta}_\mathcal{F}$, $n$ independent $q_T$-samples}
	~\\
	Initialize $i=1,\dots,n + \nabla n: \bm{x}_i^0 \sim p_0$ and $\bm{q}_i^0 = p_0(\bm{x}_i^0)$\\
	
	\While{$\widehat{\delta}_w > \delta_{\mathrm{thresh}}$}{
		Propagate all particles ($i=1,\dots,n+n_\nabla$):\\
		$\bm{x}_i^{t+1} = \bm{x}_i^t + \epsilon_t \widehat{\phi}_t^*(\bm{x}_i^t)$   with  
		$ \widehat{\phi}_t^*(\bm{y}) = \frac{1}{n} \sum\limits_{i=1}^{n_{\nabla}} k(\bm{x}_i^t,\bm{y}) [\nabla_{\bm{x}} \log F(\bm{x}_i^t) + \nabla_{\bm{x}} \log p_0(\bm{x}_i^t)] + \nabla_{\bm{x}} k(\bm{x}_i^t,\bm{y})$
		
		Track the PDF of all independent particles ($i=n_{\nabla} + 1,\dots,n_{\nabla} + n$):\\
		$q(\bm{x}_i^t) = q\left(\bm{x}_i^{t-1}\right) \left|\mathrm{det} \nabla \mathcal{T}_t(\bm{x}_i^{t-1})\right|^{-1}$
	}
	\vspace{.1cm}

	Estimate $\widehat{p}_{\mathcal{F}}^{\text{SVRE}}$ and $\widehat{\delta}_{\mathcal{F}}^{\text{SVRE}}$ according to \cref{eq:phat_svgd} and \cref{eq:deltahat_svgd} and \textbf{return}\\
\end{algorithm2e}

\section{Numerical examples}
\label{sec:numerical}

\subsection{Error measures \& algorithm settings}
\label{sec:settings}
We quantify the accuracy of rare event probability estimates using the relative root-mean-squared error (relative RMSE):
\begin{equation}
	\label{eq:rRMSE}
	\text{rRMSE} = \frac{1}{p_{\mathcal{F}}} \sqrt{\mathbb{E}[(p_{\mathcal{F}} - \widehat{p}_{\mathcal{F}})^2]} =  \frac{1}{p_{\mathcal{F}}} \sqrt{\smash[b]{\underbrace{(p_{\mathcal{F}} - \mathbb{E}[\widehat{p}_{\mathcal{F}}])^2}_{\text{bias}^2} + \underbrace{ \mathbb{V}[ \widehat{p}_{\mathcal{F}}]}_{\text{variance}}}}
 =   \sqrt{\smash[b]{\underbrace{\left( \frac{p_{\mathcal{F}} - \mathbb{E}[\widehat{p}_{\mathcal{F}}]}{p_{\mathcal{F}}}\right)^2}_{\text{rel. bias}^2} + \underbrace{	\vphantom{\left( \frac{p_{\mathcal{F}} - \mathbb{E}[\widehat{p}_{\mathcal{F}}]}{p_{\mathcal{F}}}\right)^2} \frac{ \mathbb{V}[ \widehat{p}_{\mathcal{F}}]}{p_{\mathcal{F}}^2}}_{\text{rel. variance}}}}.
\end{equation}
~\\~\\
We compare the relative RMSE produced by rare event-SVGD with three state-of-the-art gradient-based rare event simulation methods including HMC-based subset simulation (HMC-SS) \cite{Wang2019b} 
and two gradient-based importance sampling approaches referred to as ASTPA \cite{Papakonstantinou2023} and iCEred \cite{Uribe2020}, the former of which is also based on HMC. In these references as well as for SVRE, 500 independent randomly initialized runs of the methods are used to estimate the expectation in \cref{eq:rRMSE} and we do the same for SVRE.
Unless noted otherwise, we run the SVRE algorithm with $n_{\nabla} = 20$, $\delta_{\mathrm{thresh}} = 5$ and $\bar{\epsilon} = 0.1 \dots 1$ depending on the choice of normalization scheme for $\epsilon_t$. The displayed results for HMC-SS \cite{Wang2019b}, ASTPA \cite{Papakonstantinou2023} and iCEred \cite{Uribe2020} are taken from the original references.
\\~\\
In general, SVRE may be run with arbitrary absolutely continuous input distribution $p_0$.
For the numerical experiments in this work, the input random vector $\bm{X}$ is transformed to standard-normal space such that $p_0$ is the standard-normal PDF. This facilitates a unified implementation of SVRE for arbitrary input distributions (given an isoprobabilistic transformation from $\mathcal{X}$ to standard-normal space) and choosing non-adaptive kernel bandwidths in a problem-independent manner. We draw initial samples from the standard-normal $p_0$ by transforming a low-discrepancy (Sobol') sequence \cite{Niederreiter1992}. Technically, samples generated with low-discrepancy sequences are not independent, which may introduce bias in the failure estimate \cref{eq:phat_svgd} and render the estimator coeffcient of variation formula \cref{eq:deltahat_svgd} invalid. We find these effects are negligible in all investigated examples, however, in order to preserve sample independence one could replace the low-discrepancy series with a latin hypercube sampling step.
\\~\\
With the exception of the random initial sample, SVRE is a deterministic algorithm (as is SVGD). It has been pointed out that samples generated with deterministic particle-based inference algorithms are more prone to collapse to local modes under certain conditions compared to fully stochastic approaches such as MCMC or stochastic gradient descent methods \cite{Welling2011, Dangelo2021, Zhang2020}. 
In our experiments, this concerns a relatively small percentage of all 500 runs (less than 5\% in any experiment). These cases can be identified reliably as they are associated with large estimator coefficients of variation. Therefore, we exclude these runs ($\widehat{\delta}^{\mathrm{SVRE}}_{\mathcal{F}} > 0.5$) from relative RMSE computations. 
\subsection{Test functions}
\label{sec:test_functions}
\subsubsection{Linear LSF}
\label{sec:linear_lsf}
\begin{figure}[h!]
	\centering
	\subfloat[][]{\includegraphics[width=0.33\textwidth]{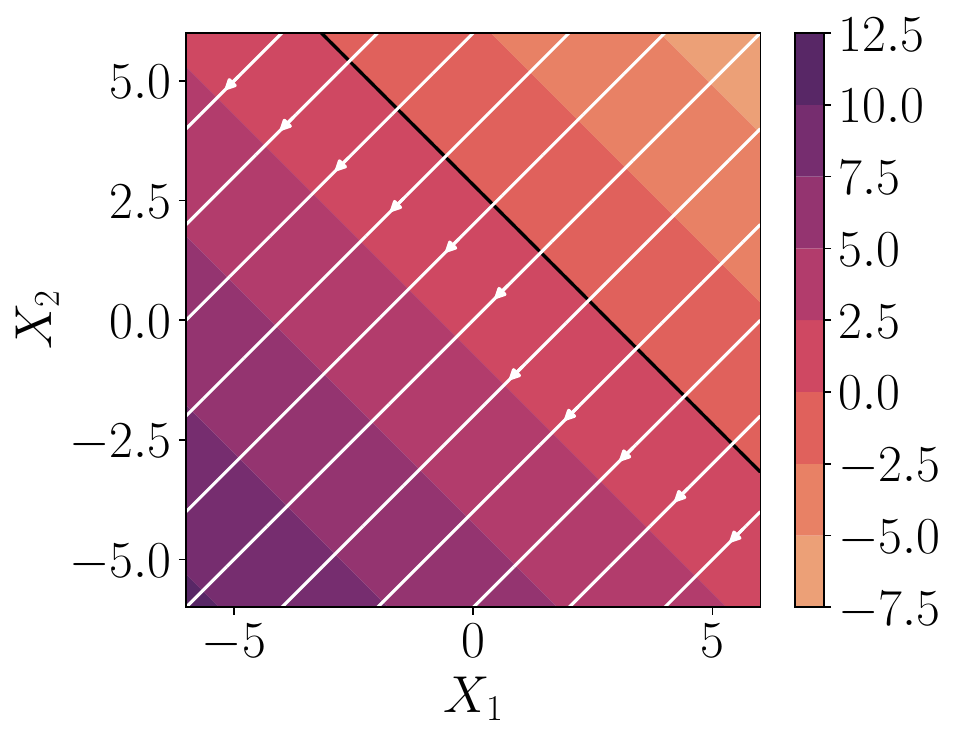}} \hfill
	\subfloat[][]{\includegraphics[width=0.33\textwidth]{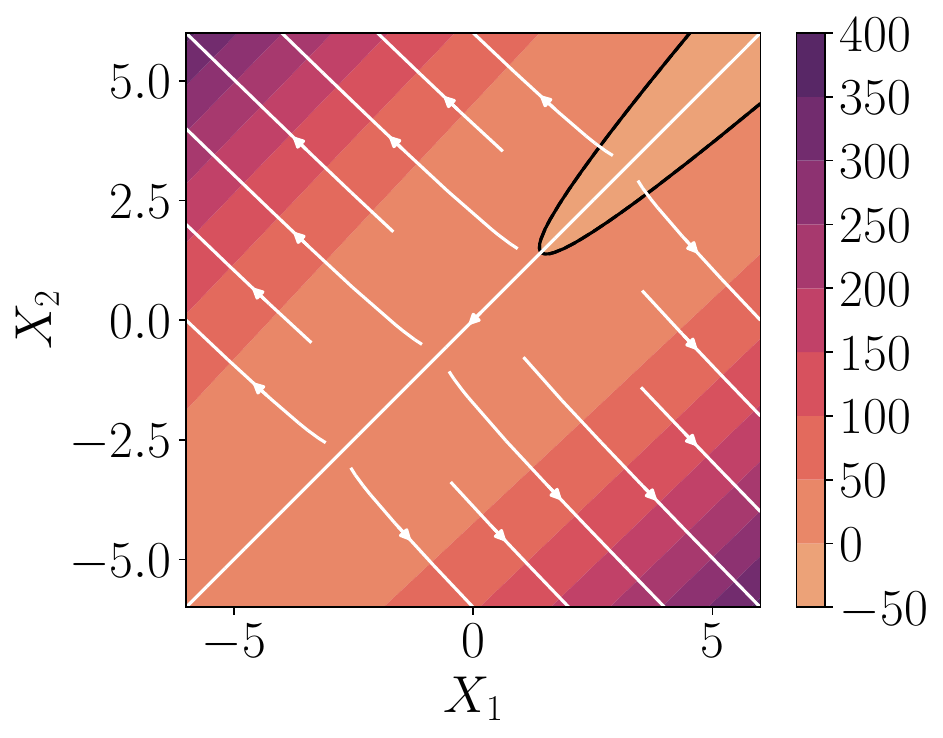}} \hfill
	\subfloat[][]{\includegraphics[width=0.33\textwidth]{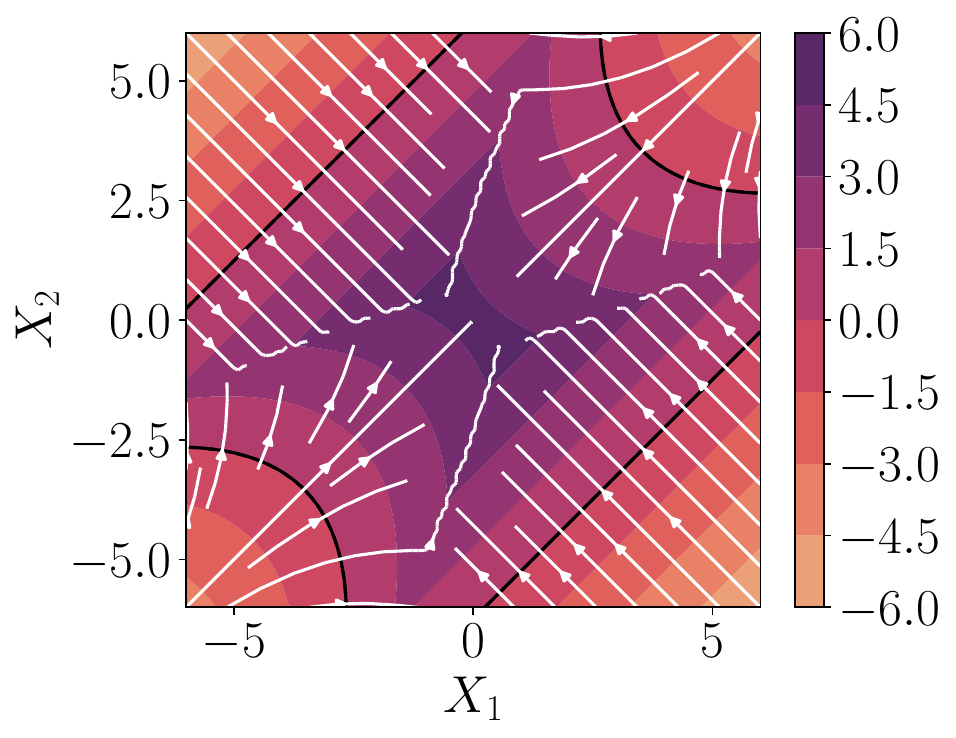}}
	\caption{Contour plots of the three considered test LSFs with failure hypersurface $G(\bm{X})=0$ (black line) and streamlines of the LSFs (white lines). Left: linear LSF with $d=2$, center: quadratic LSF with $d=2$, right: four branch LSF.}
	\label{fig:lsf_plots}
\end{figure}

The following linear limit-state function is a standard testbed example from reliability analysis and rare event simulation \cite{Engelund1993} (see \cref{fig:lsf_plots}, left). It reads
\begin{equation*}
	g(\bm{x}) = \beta - \frac{1}{\sqrt{d}} \sum\limits_{i=1}^d x_i,~~\nabla g(\bm{x}) = - \frac{1}{\sqrt{d}} \bm{1},~~\bm{x} \sim \mathcal{N}(\bm{0},\mathbf{I})
\end{equation*}
The exact rare event probability is available in closed form as $p_\mathcal{F} = \Phi(- \beta)$ where $\Phi(\cdot)$ is the cumulative distribution function of a standard-normal random variable. The reference solution is dimension-independent and the problem lends itself well to investigate the performance of rare event simulation algorithms in a simple setting across varying problem dimension with prescribed target probability level.
\setlength\extrarowheight{-2pt}
\begin{table*}[h!]
	\setlength{\tabcolsep}{0.1cm}
	\centering
	\caption{Error (rRMSE) and cost (required number of model and model gradient evaluations) comparison of $p_{\mathcal{F}}$-estimates by HMC-SS \cite{Wang2019b}, ASTPA \cite{Papakonstantinou2023} and our method for the linear LSF with $d=100$. SVRE parameters are chosen as $n_{\nabla} = 20$, $cv_{\mathrm{thresh}} = 5$, $\bar{\epsilon} = 1$ with $\ell_2$ normalization. For \cite{Wang2019b,Papakonstantinou2023}, the number of model calls equals the number of gradient calls.}
	\label{tab:linear_results}
	\begin{tabular}{cc|ccc|ccc}
		\hline
		 & LSF parameter& \multicolumn{3}{c|} {rRMSE} & \multicolumn{3}{c} {gradient (+ model) calls}  \\ 
		& (exact failure probability) & ASTPA & HMC-SS & SVRE &  ASTPA & HMC-SS & SVRE \\ 
		\hline
		\multirow{4}{*}{\textbf{Linear}}&$\beta = 4$ ($p_{\mathcal{F}} = 3.17 \cdot 10 ^{-5}$) & 0.29 & 0.21 & 0.08 & 5617 (+5617) & 4600 (+4600)  & 72 (+1000) \\
		\cline{2-8}
		&$\beta = 5$ ($p_{\mathcal{F}} =2.87 \cdot 10 ^{-7}$)& 0.30 & 0.28 & 0.10 & 6641 (+6641) & 6400 (+6400) & 93 (+1000) \\
		\cline{2-8}
		&$\beta = 6$ ($p_{\mathcal{F}} =0.99 \cdot 10 ^{-9}$) & 0.23 & 0.35 & 0.11 & 7635 (+7635) & 8731 (+8731) & 112 (+1000) \\
		\cline{2-8}
		&$\beta = 7$ ($p_{\mathcal{F}} =1.28 \cdot 10 ^{-12}$) & 0.29 & - & 0.11 & 8639 (+8639)  & -  & 132 (+1000)\\
		\hline
	\end{tabular}
\end{table*}
\cref{tab:linear_results} shows a comparison of relative RMSE and total number of LSF and LSF gradient evaluations produced by the two HMC-based reference methods and our SVRE algorithm at different target probability levels between $\mathcal{O}(10^{-5})$ and $\mathcal{O}(10^{-12})$ and for $d=100$. The relative RMSE of HMC-SS \cite{Wang2019b} increases with decreasing target probability where as ASTPA \cite{Papakonstantinou2023} and SVRE errors are independent of the target probability magnitude. At around 10 \%, the rRMSE of SVRE is noticeably smaller than that of the other two methods, while generating a speedup of two orders of magnitude (in terms of LSF gradient evaluations). 

\subsubsection{Quadratic LSF}
Adding a quadratic term  to the linear LSF from \cref{sec:linear_lsf} in the $x_1$-$x_2$-plane leads to the following quadratic LSF (see \cref{fig:lsf_plots}, center), which was first presented in \cite{Papaioannou2015}:
\begin{equation*}
	g(\bm{x}) = \beta + \frac{\kappa}{4}(x_1 - x_2)^2 - \frac{1}{\sqrt{d}} \sum\limits_{i=1}^d x_i,~
	\nabla g(\bm{x}) = \left[  \frac{\kappa}{2}(x_1 - x_2) - 1/\sqrt{d}, -  \frac{\kappa}{2}(x_1 - x_2) - 1/\sqrt{d}, -\frac{1}{\sqrt{d}} \bm{1}\right],
\end{equation*}
where $\bm{x} \sim \mathcal(N)(\mathbf{0},\mathbf{I})$. $\kappa$ governs the curvature of the LSF at the most probable point of failure and larger $\kappa$ implies larger curvature, which in turn implies smaller $p_{\mathcal{F}}$. Typical choices are $\kappa = -1 \dots 10$ \cite{Wang2019b}, where in this work we present results for the highest curvature setting at $\kappa = 10$. The exact failure probability cooresponding to $\kappa = 10$ is $p_{\mathrm{F}} = 4.73 \cdot 10^{-6}$ independent of the problem dimension.
\begin{table*}[h!]
	\setlength{\tabcolsep}{0.1cm}
	\centering
	\caption{Error (rRMSE) and cost(required number of model and model gradient evaluations) comparison of  $p_{\mathcal{F}}$-estimates by HMC-SS \cite{Wang2019b}, ASTPA for rare events \cite{Papakonstantinou2023} and our method for the quadratic LSF. SVRE parameters are chosen as $n_{\nabla} = 20$, $cv_{\mathrm{thresh}} = 5$ and $\bar{\epsilon} = 0.1$ for RMSProp and $\bar{\epsilon} = 1$ for $\ell_2$ normalization. For \cite{Wang2019b,Papakonstantinou2023}, the number of model calls equals the number of gradient calls.}
	\label{tab:quadratic_results}
	\begin{tabular}{cc|ccccc|ccccc}
		\hline
		&& \multicolumn{5}{c|} {rRMSE} & \multicolumn{5}{c} {gradient (+ model) calls}  \\ 
		&dimension &  ASTPA & HMC-SS & SVRE  & SVRE & SVRE & ASTPA & HMC-SS & SVRE & SVRE  & SVRE  \\ 
		& & & & (RMSProp) & (adapt $\bar{\epsilon}$)  &($\ell_2$) & & &  (RMSProp) & (adapt $\bar{\epsilon}$) & ($\ell_2$) \\
		\hline
		\multirow{4}{*}{\rotatebox{90}{\textbf{Quadratic}}} & \multirow{2}{*}{$d = 2$} & \multirow{2}{*}{0.18} & \multirow{2}{*}{-} & \multirow{2}{*}{0.14} & \multirow{2}{*}{0.12} & \multirow{2}{*}{0.11} & 573  & - & 443 & 202 & 356 \\
		& & &  & &  &  & (+573)&  & (+$10^3$)& (+$10^3$) & (+$10^3$) \\
		
		\cline{2-12}
		&\multirow{2}{*}{$d = 100$} & \multirow{2}{*}{-} & \multirow{2}{*}{0.31} & \multirow{2}{*}{0.18} & \multirow{2}{*}{0.14} & \multirow{2}{*}{0.20} &  - & 19747  & 88 & 86  & 341 \\
		&  &   & & & & & & (+19747)  &  (+$10^3$)&  (+$10^3$) &  (+$10^3$) \\
		\hline
	\end{tabular}
\end{table*}
\cref{tab:quadratic_results} summarizes the computational cost and accuracy of SVRE as obtained from our numerical experiments as well as the performance of the two HMC-based reference methods. We compare three different settings of SVRE, among them two settings using RMSProp normalization and either a constant base learning rate of $\bar{\epsilon} = 0.1$ or an adaptive base learning  rate as well as one setting using $\ell_2$ normalization. Reference results are available in the literature from ASTPA at $d=2$ and from HMC-SS at $d = 100$.
\\~\\
The rRMSE of all three SVRE versions is between 10 \% and 20 \% and consistently lower than the rRMSE of the reference methods in both dimensional settings. The required number of gradient evaluations is slightly lower compared to ASTPA in $d=2$ and significantly (by more than two orders of magnitude) lower than that of HMC-SS in $d=100$. Interestingly, SVRE requires less steps to move samples to the rare event in the high-dimensional setting and hence, the total number of LSF gradients required is smaller in that setting. We suspect that this behaviour occurs in highly degenerate examples (a similar behavior can be observed for the linear example) due to a concentration-of-measure-effect that reduces the variance contributed by the second term in the LSF. In the low-dimensional setting, gradient evaluations are reduced by more than half when the base learning rate $\bar{\epsilon}$ is chosen adaptively according to \cref{appendix:B}.
\subsubsection{Four branch LSF}
The following four branch LSF is useful to test the SVRE performance on problems with several (here: four) relevant failure domains,  (see \cref{fig:lsf_plots}, right):
\begin{align*}
	g(\bm{x}) &= \gamma - \min 
		\begin{cases}
			 g_1 = 0.1(x_1-x_2)^2 - (x_1+x_2)/\sqrt{2} + 3 \\
			 g_2 = 0.1(x_1-x_2)^2 + (x_1+x_2)/\sqrt{2} + 3 \\
			 g_3 = x_1 - x_2 + 7/\sqrt{2} \\
			 g_4 = x_2 - x_1 + 7/\sqrt{2}
		\end{cases}\\
	\nabla g(\bm{x}) &=  - 
		\begin{cases}
			[0.2(x_1-x_2) - 1/\sqrt{2}, -0.2(x_1-x_2) - 1/\sqrt{2}], & \argmin\limits_{1,2,3,4} g_i = 1\\
			[0.2(x_1-x_2) + 1/\sqrt{2}, -0.2(x_1-x_2) + 1/\sqrt{2}], &\argmin\limits_{1,2,3,4} g_i = 2\\
			[1 + 7/\sqrt{2}, -1 + 7/\sqrt{2} ], & \argmin\limits_{1,2,3,4} g_i = 3\\
			[-1 + 7/\sqrt{2}, 1 + 7/\sqrt{2} ], & \argmin\limits_{1,2,3,4} g_i = 4
		\end{cases},
\end{align*}
where $\bm{x} \sim \mathcal{N}(\bm{0},\mathbf{I})$ and the target rare event probability depends on $\gamma$. We investigate $\gamma = 0,2,4$, which correspond to probabilities between $4.45 \cdot 10^{-3}$ and $5.57 \cdot 10^{-9}$.
\begin{table*}[h!]
	\setlength{\tabcolsep}{0.1cm}
	\centering
	\caption{Error (rRMSE) and cost (required number of model and model gradient evaluations) comparison of $p_{\mathcal{F}}$-estimates by ASTPA \cite{Papakonstantinou2023} and our method for the four branch LSF. SVRE parameters are chosen as $\delta_{\mathrm{thresh}} = 5$, $\bar{\epsilon} = 0.25$ and RMPSprop normalization. For \cite{Wang2019b,Papakonstantinou2023}, the number of model calls equals the number of gradient calls.}
	\label{tab:fourbranch_results}
	\begin{tabular}{cc|ccc|ccc}
		\hline
		&LSF & \multicolumn{3}{c|} {rRMSE} & \multicolumn{3}{c} {gradient (+ model) calls}  \\ 
		& parameter &  ASTPA & \multicolumn{2}{c|} {SVRE} & ASTPA & \multicolumn{2}{c} {SVRE}  \\
		& ($p_{\mathcal{F}} $)& &$n_{\nabla} = 50$ & $n_{\nabla} = 100$ & &$n_{\nabla} = 50$ & $n_{\nabla} = 100$ \\
		\hline
		&$\gamma = 0$ & \multirow{2}{*}{0.18} & \multirow{2}{*}{0.13}  & \multirow{2}{*}{0.13} & 573 &  170.2 & 357.8 \\
		&  ($ 4.451 \cdot 10 ^{-3}$) &   &   &   & (+ 573)  &   (+$10^3$)& (+$10^3$) \\
		\cline{2-8}
		\textbf{Four}&$\gamma = 2$ & \multirow{2}{*}{-} & \multirow{2}{*}{0.28} & \multirow{2}{*}{0.27} &  \multirow{2}{*}{-} &  319.3 & 681.2  \\
		\textbf{Branch}&  ($9.620 \cdot 10 ^{-6}$)  & & & & & (+$10^3$)& (+$10^3$) \\
		\cline{2-8}
		&$\gamma = 4$  & \multirow{2}{*}{-} & \multirow{2}{*}{0.29} & \multirow{2}{*}{0.25} &  \multirow{2}{*}{-} &    465.8& 1026.3 \\
		& ($ 5.596 \cdot 10 ^{-9}$) & &   &  &  &  (+$10^3$)&  (+$10^3$) \\
		\hline
	\end{tabular}
\end{table*}
\cref{tab:fourbranch_results} shows the performance of the SVRE estimator using $50$ and $100$ gradient samples per iteration 
and compares these to ASTPA results stated in \cite{Papakonstantinou2023} for the case $\beta = 0$ (no reference results are available for larger $\beta$). The SVRE estimators at both $n_\nabla$ settings slightly outperform ASTPA in terms of relative RMSE using approximately one third of ASTPA's total gradient evaluations.
Increasing $\beta = 2, 4$ leads to slightly elevated rRMSE and an increased number of iterations to drive samples to the failure domain resulting in an increased number of total gradient evaluations for SVRE and both $n_\nabla$-choices. 
Increasing $n_\nabla$ from $50$ to $100$ reduces the rRMSE marginally and -- as expected -- roughly doubles the total number of gradient evaluations. In \cite{Zhang2020}, it is shown that there exists an optimal number of samples for SVGD implying that increasing the number of samples need not necessarily improve the performance of the algorithm. Within SVRE, this statement pertains to inducing samples ($n_\nabla$), while increasing the number of estimation samples ($n$) will always improve the performance of SVRE, i.e., it will always decrease the estimator's rRMSE.
\subsection{Engineering examples}
\label{sec:engineering_examples}

\subsubsection{Darcy flow across porous aquifer}
\label{sec:darcy}
\begin{figure}[h!]
	\centering
	\subfloat[][]{\includegraphics[width=0.49\textwidth]{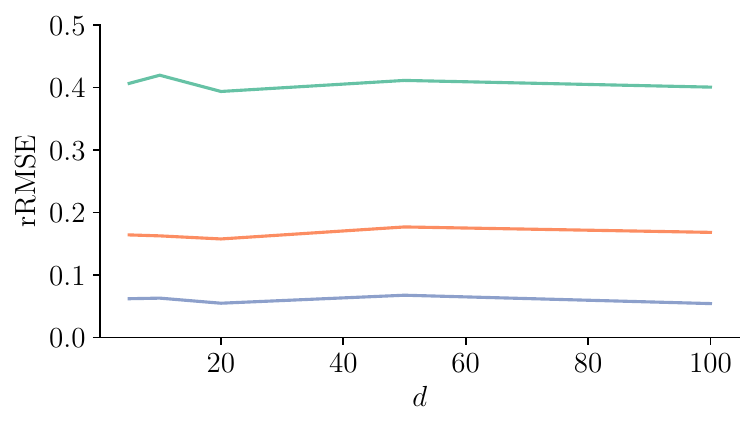}} \hfill
	\subfloat[][]{\includegraphics[width=0.49\textwidth]{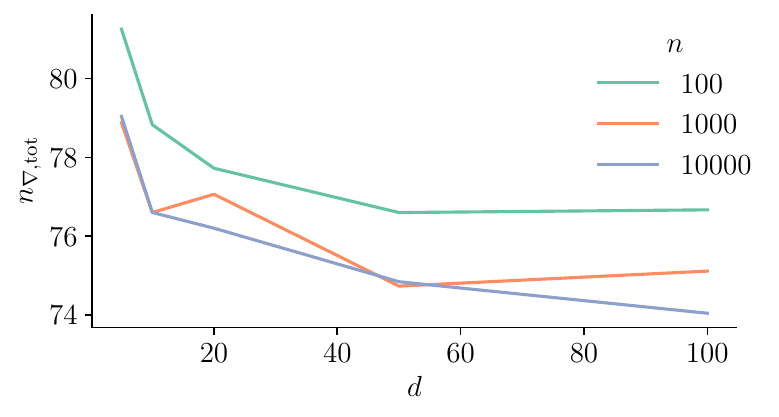}}
	\caption{Parameter study: SVRE accuracy (left) and cost (right) across different problem dimensions $d$ while varying the number of samples used for estimating $p_{\mathcal{F}}$. SVRE parameters are chosen as $n_{\nabla} = 20$, $cv_{\mathrm{thresh}} = 5$, $\bar{\epsilon} = 1$ with $\ell_2$ normalization.}
	\label{fig:diffusion_results}
\end{figure}

\paragraph{Problem description}
We consider the following 1-D boundary value problem on $D = [0,1]$
\begin{equation}
	y \in D:~~\frac{\mathrm{d}}{\mathrm{d} y} \left( \kappa(y, \omega) \frac{\mathrm{d} u(y,\omega)}{\mathrm{d} y} \right) = - J(y), ~ \frac{\mathrm{d}  u(y,\omega)}{\mathrm{d} y}\Big|_{y=0} = - F(\omega), ~u(1, \omega) = 1,
\end{equation}
where $u(y, \omega)$ describes the pressure head resulting from the stochastic diffusivity field $\kappa$ as well as the given Dirichlet ($y=1$) and Neumann ($y=0$) boundary conditions. $F(\omega)$ represents a random flux boundary condition at the left boundary of $D$ and is modelled as a Gaussian with mean $\mu_F = 2$ and variance $\sigma^2_F = 0.5$. The diffusivity $\kappa$ is modelled as a log-normal random field (RF), which is generated by discretizing the underlying Gaussian RF $\ln \kappa$ with mean $\mu_{\ln \kappa} = 1$, variance $\sigma^2_{\ln \kappa} =0.3$ and correlation kernel $k(y- y') = \exp\{ - |y - y'|/\ell\}~(\ell = 0.1)$ using a Karhunen-Lo{\`e}ve-expansion with $d-1$ terms:
\begin{equation}
	\label{eq:KL_kappa}
	\kappa(y) = \exp\left\{\mu_{\ln \kappa}  + \sigma_{\ln \kappa} \sum_{i=1}^{d-1} \sqrt{\lambda^\kappa_i} \varphi^\kappa_i(y) \xi_i\right\}.
\end{equation}
$\{\lambda^\kappa_i, \varphi^\kappa_i(y)\}_{i=1}^{d-1}$ are the first $d-1$ eigenpairs of the correlation kernel where the eigenvalues are ordered in descending order and $\bm{\xi}$ is a $(d-1)$-dimensional standard-normal random vector. The source term $J$ is modelled as a mixture of four Gaussian plumes located equidstantly across $D$, i.e.,
\begin{equation}
	J(y) = 0.8 \sum_{i=1}^4 \mathcal{N}(y | \mu_i = 0.2 \cdot i, \sigma_i = 0.05).
\end{equation}
We are interested in the maximum pressure head exceeding a critical value $p_{\mathrm{thresh}} = 2.7$, hence the LSF is defined as
\begin{equation}
	\label{eq:diffusion_lsf}
	g(\bm{X}) = p_{\mathrm{thresh}}  - \max(u(y,\bm{X})),
\end{equation}
where $\bm{X} \vcentcolon = [F,\xi_1,\dots,\xi_{d-1}]$.
This example is taken from and discussed in greater detail in \cite{Uribe2020b}.
\paragraph{Results}
\cref{fig:diffusion_results} shows a comparison of accuracy and cost for SVRE estimates of the maximum pressure exceedance probability resulting from the LSF \cref{eq:diffusion_lsf} for problem dimensions $d=[5,10,20,50,100]$. The exceedance probability estimate increases with increasing $d$ , i.e., increasing number of KLE terms in \cref{eq:KL_kappa}, yet remains on the order $\mathcal{O}(10^{-5})$ in all investigated cases \cite{Uribe2020b}.
The relative RMSE of all SVRE estimates decay approximately with rate $n^{-1/2}$, which is expected as the sample split described in \cref{sec:sample_split} yields $n$ independent samples for the IS estimate. Further, the error remains constant as $d$ increases. Likewise, the total number of gradient evaluations is insensitive to the problem dimension $d$ as well as the number of estimation samples $n$. The total cost incurred of course still increases with $n$ as it is equal to $n_{\nabla,\mathrm{tot}} + n$.
\begin{figure}[h!]
	\centering
	\subfloat{\includegraphics[width=\textwidth]{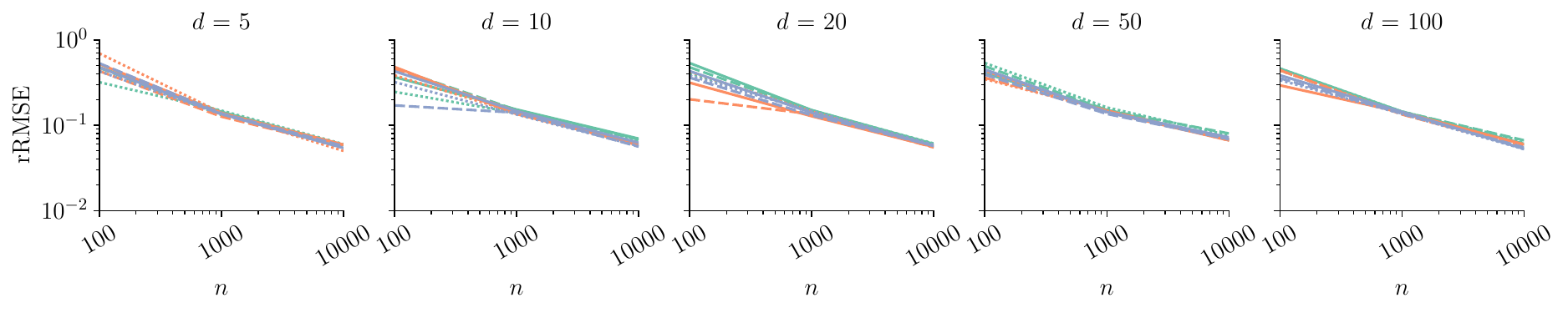}}\\ \vspace{-.25cm}
	\subfloat{\includegraphics[width=\textwidth]{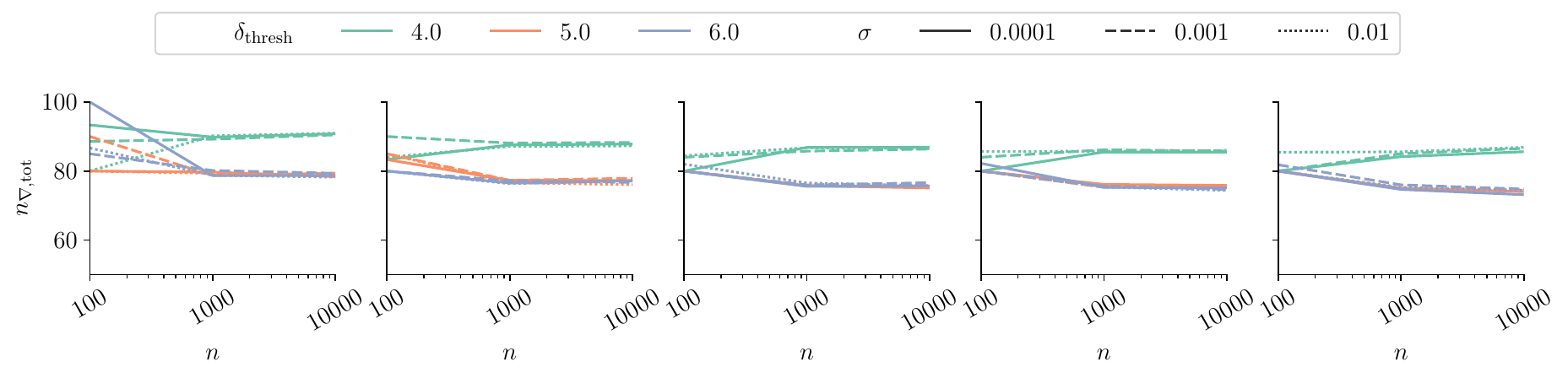}}
	\caption{Diffusion problem parameter study: relative RMSE (top) and cost in terms of total number of gradient evaluations per run (bottom) for various termination criteria $\delta_{\mathrm{thresh}}$ and degrees of smoothing $\sigma$.}
	\label{fig:diffusion_parameter}
\end{figure}
\cref{fig:diffusion_parameter} shows SVRE accuracy and cost for varying parameters $\sigma$ (the degree of LSF smoothing) and the exit criterion $\delta_{\mathrm{thresh}}$. Both cost and accuracy of the SVRE estimator are insensitive to the choice of $\sigma$ which is varied over two orders of magnitude ($10^{-2}\dots10^{-4}$). An intuitive finding is that choosing $\delta_{\mathrm{thresh}}$ smaller results in an increased overall number of gradient evaluations as the method is run for more iterations on average. However, the estimator's accuracy also decreases at low sample sizes if $\delta_{\mathrm{thresh}}$  is chosen too small, i.e., if the criterion for proximity of the IS distribution to the target distribution is too strict.
\subsubsection{Steel plate in plane stress}
\label{sec:plate}
\paragraph{Problem description}

We consider a modified version of the example given in \cite{Uribe2020,liu_and_liu_1993}, which consists of a low-carbon steel plate of length $0.32$~m, width $0.32$~m, thickness $t = 0.01$~m, and a hole of radius $0.02$~m located at the center. The Poisson ratio is set to $\nu = 0.29$ and the density of the plate is $\rho=7850$~kg/m$^3$. The horizontal and vertical displacements are constrained at the left edge. The plate is subjected to a random surface load that acts on the right narrow plate side. The load is modelled as a log-normal random variable with mean $\mu_q = 60$ MPa and $\sigma_q = 12$ MPa.
\begin{figure}[!h]
	\centering
	\begin{minipage}[t]{0.33\textwidth}
		\raisebox{-.25cm}{\includegraphics[width=\textwidth]{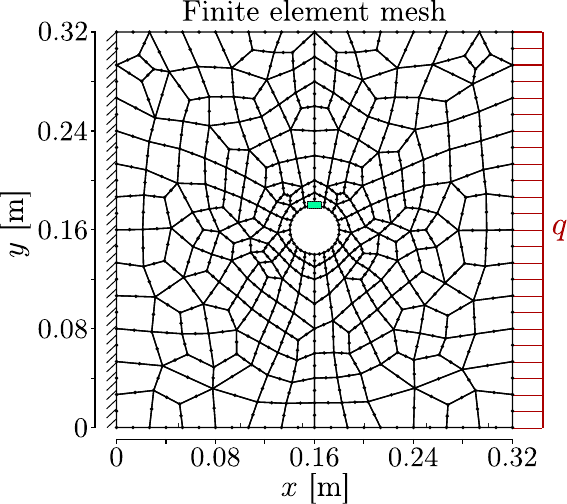}}
	\end{minipage}
	\hspace{0.1cm}
	\begin{minipage}[t]{0.35\textwidth}
	\raisebox{-.25cm}{\includegraphics[width=\textwidth]{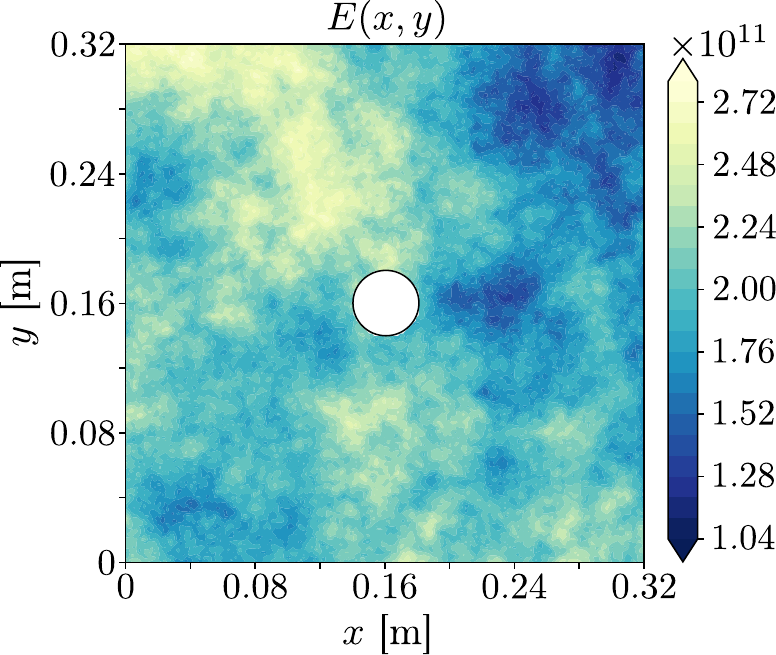}}
\end{minipage}
\hspace{0.55cm}
	\begin{minipage}[b]{0.25\textwidth}
		\subfloat{\includegraphics[width=0.33\textwidth]{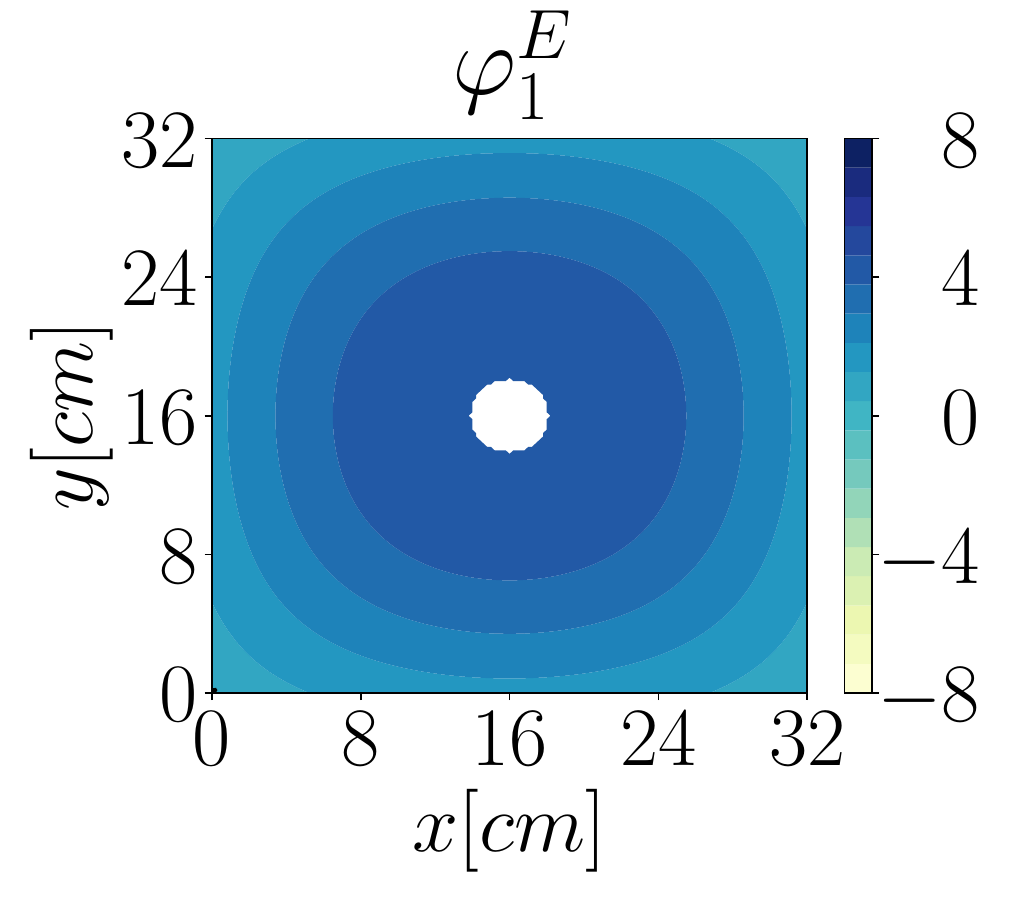}}
		\hfill 
		\subfloat{\includegraphics[width=0.33\textwidth]{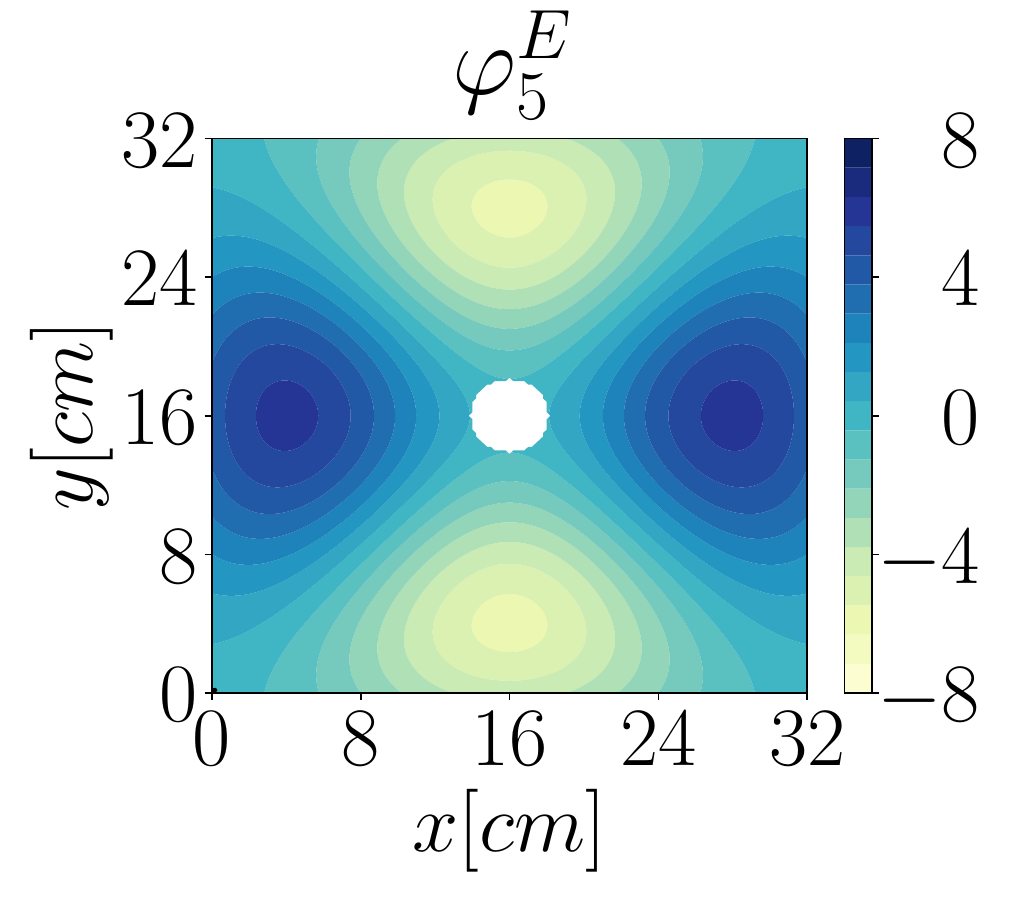}}
		\hfill
		\subfloat{\includegraphics[width=0.33\textwidth]{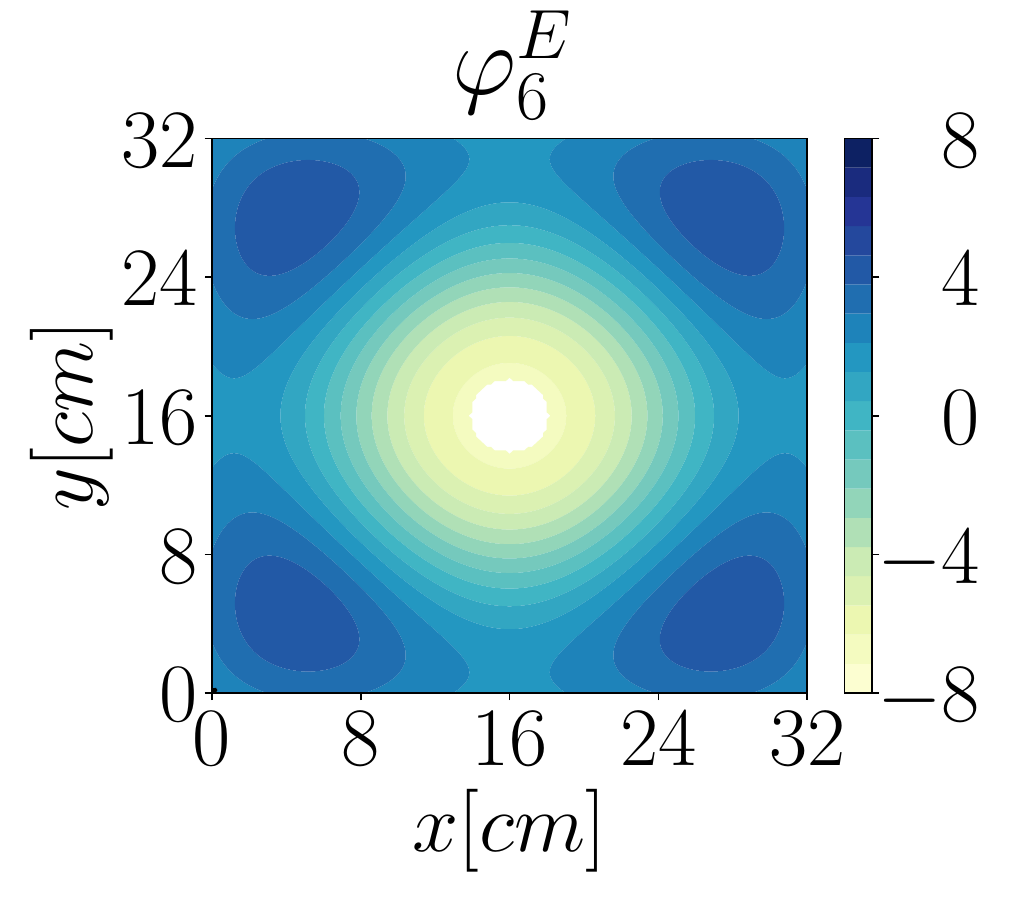}}
		\\
		\subfloat{\includegraphics[width=0.33\textwidth]{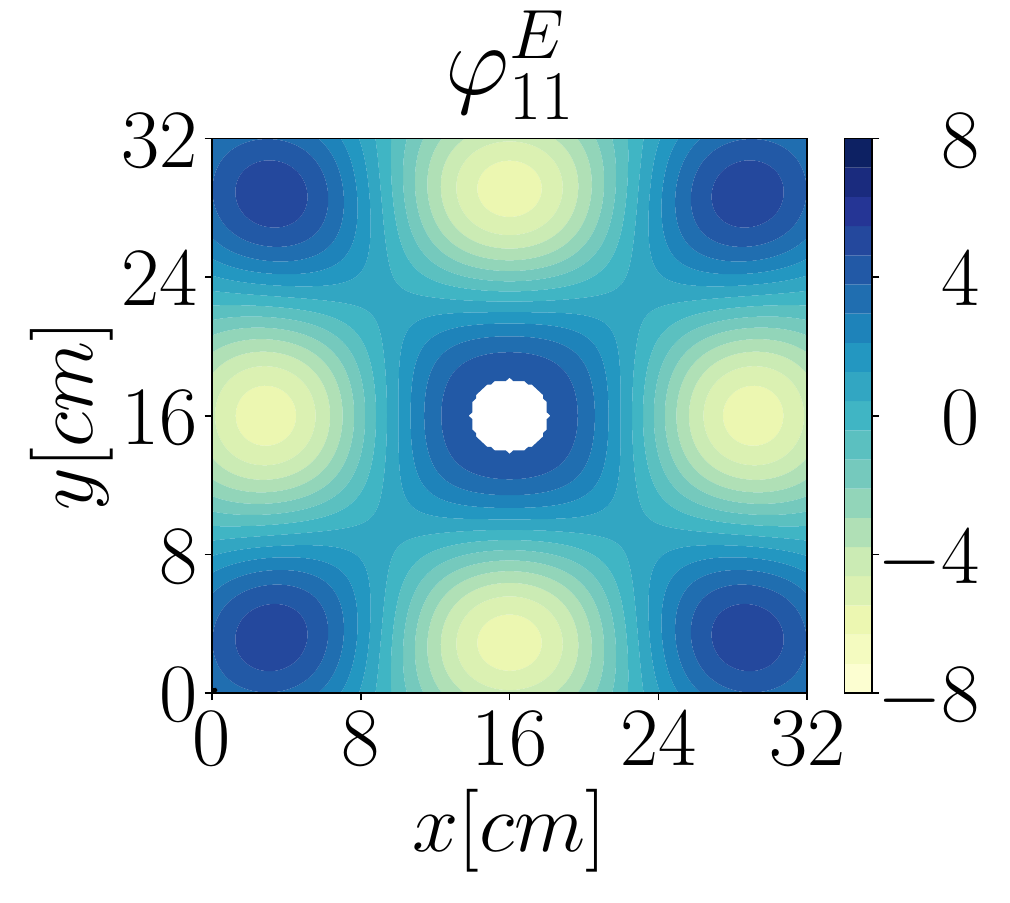}}
		\hfill 
		\subfloat{\includegraphics[width=0.33\textwidth]{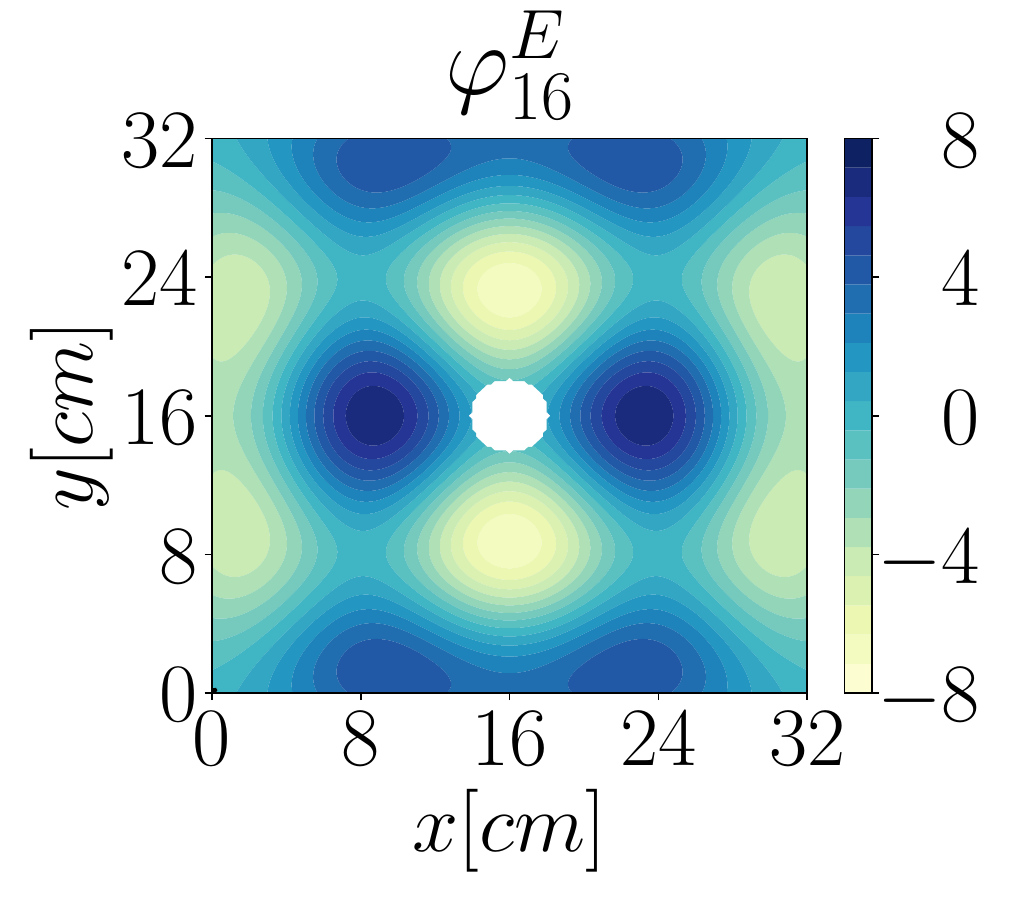}}
		\hfill
		\subfloat{\includegraphics[width=0.33\textwidth]{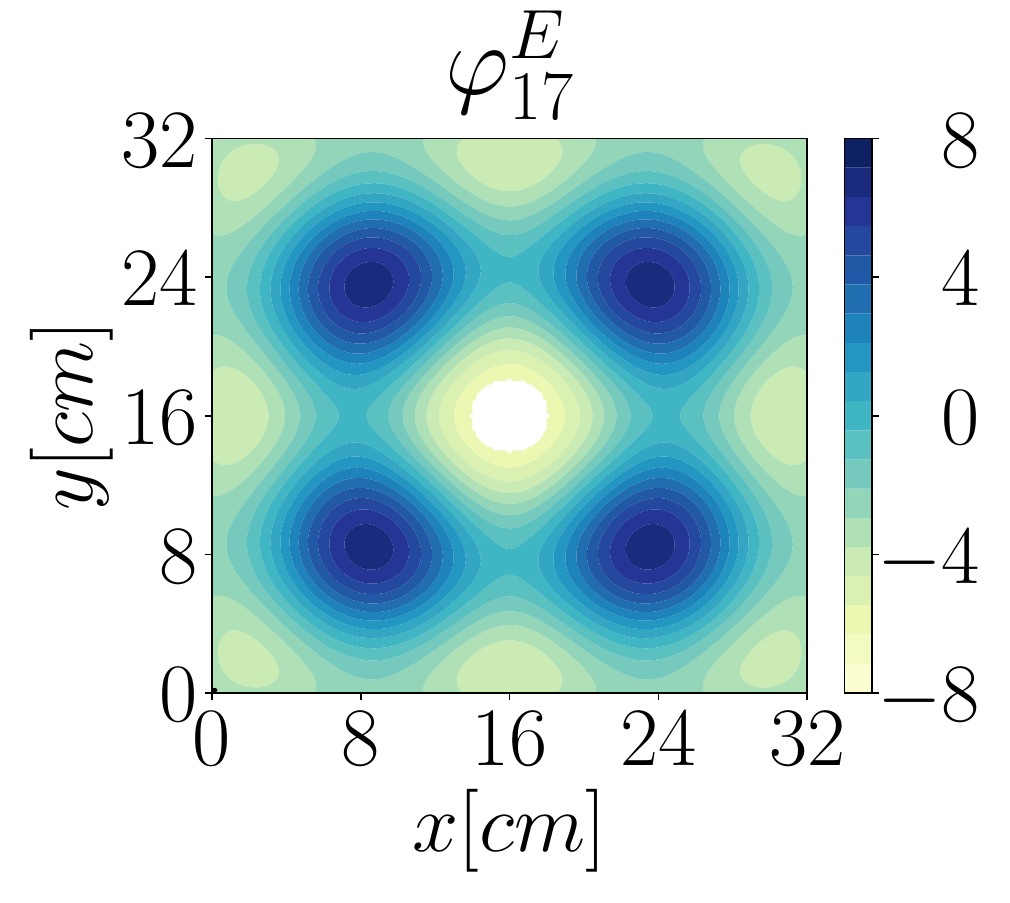}}
		\\
		\subfloat{\includegraphics[width=0.33\textwidth]{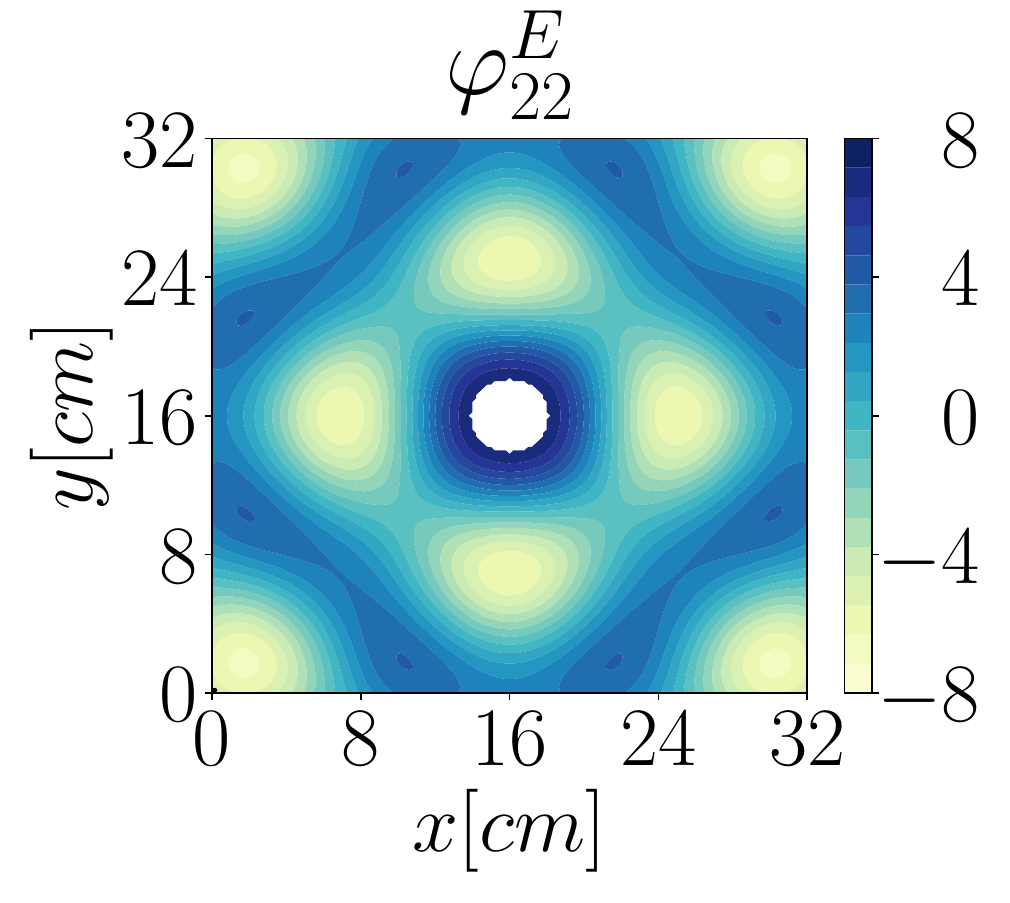}}
		\hfill 
		\subfloat{\includegraphics[width=0.33\textwidth]{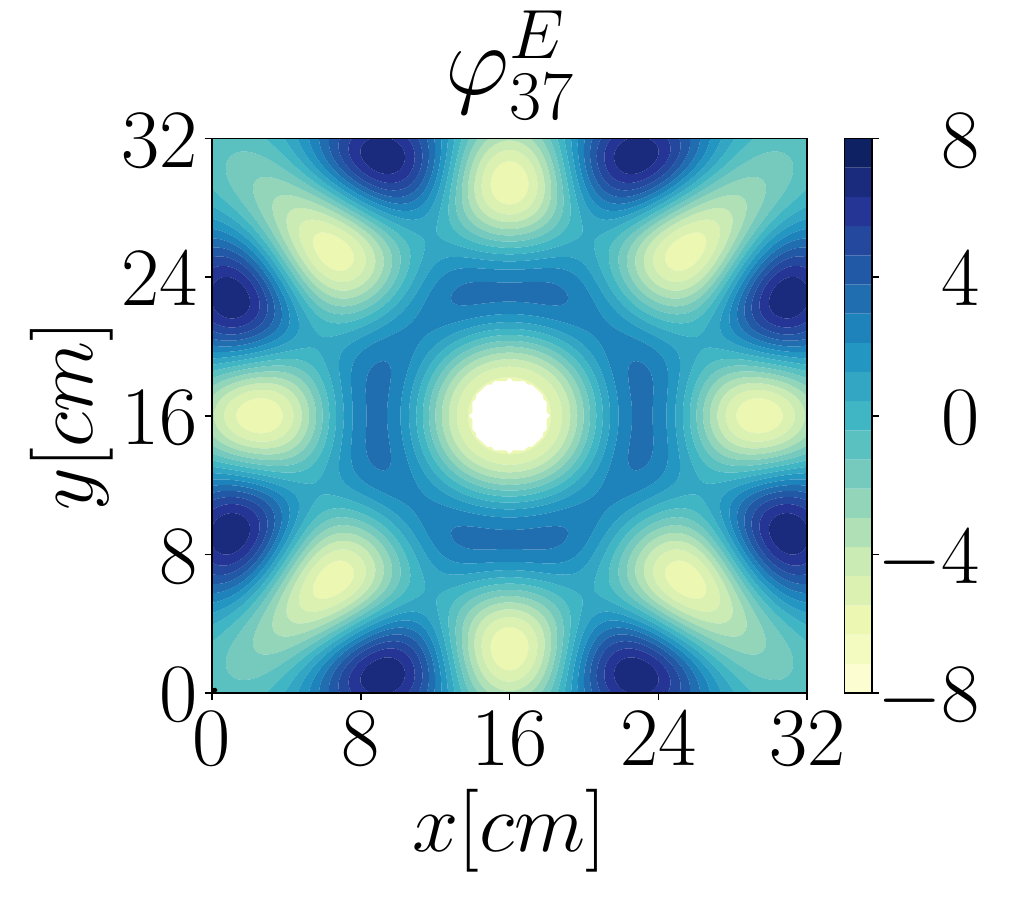}}
		\hfill
		\subfloat{\includegraphics[width=0.33\textwidth]{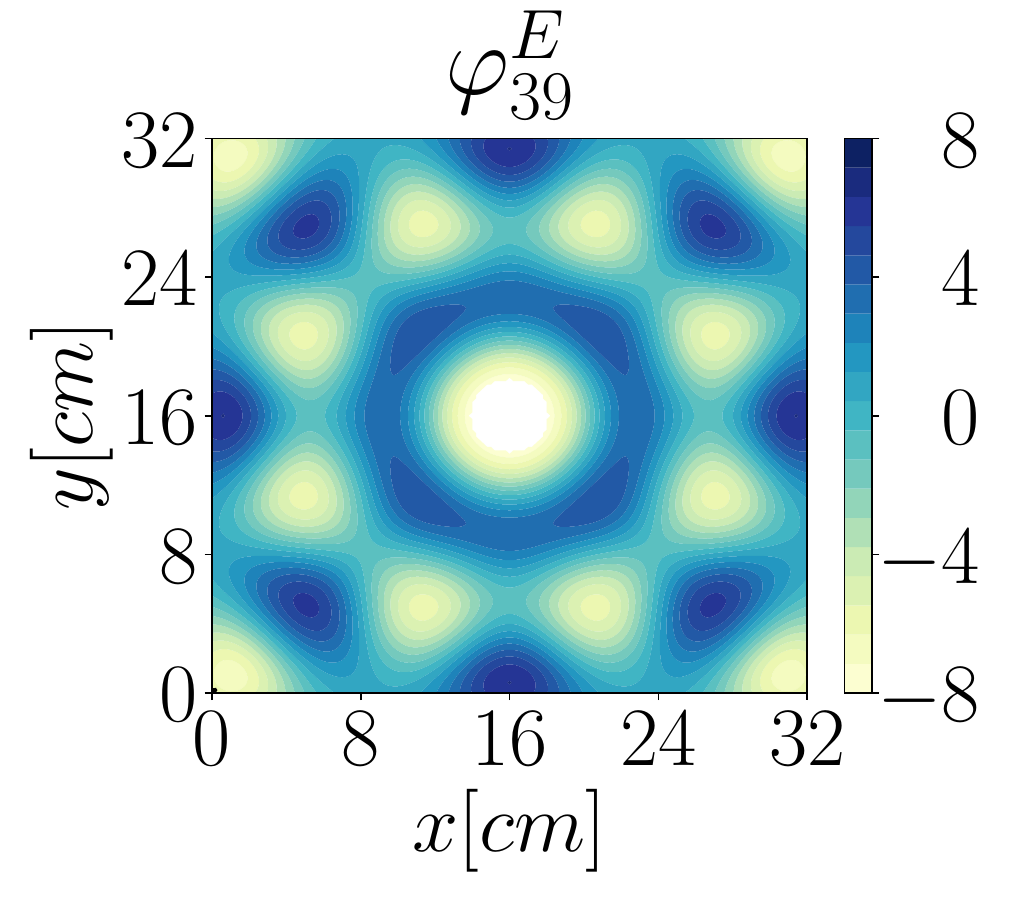}}
	\end{minipage}
	\caption{Left: FE-mesh of 2D-plate model with control node of the first principal stress $\sigma_1$. Center: realization of the Young's modulus RF. Right: eigenfunctions of the isotropic exponential covariance kernel. Taken from \cite{Uribe2020b} with permission.}
	\label{fig:plate}
\end{figure}  
The plate's Young's modulus $E(x,y)$ is considered uncertain and spatially variable. It is described by a homogeneous random field with lognormal marginal distribution, mean value $\mu_E = 2\times 10^{5}$~MPa and standard deviation $\sigma_E = 3\times 10^{4}$~MPa. 
The autocorrelation function of the underlying Gaussian field $\ln E$ is modeled by the isotropic exponential model
\begin{equation}
	\label{eq:corr_E}
	\rho_{\ln E}(\Delta x,\Delta y) = \exp \left\{-\frac{\sqrt{\Delta x^2 + \Delta y^2}}{l_E}\right\}
\end{equation}
with correlation length $l_{\ln E} = 0.04 \text{m}$.
The Gaussian random field $\ln E$ is discretized by a Karhunen-Lo\`eve-expansion with variable $d_E$ and hence variable problem dimensions (with a focus on $d_E=868$, which yields a mean error variance of 7.5\% for the discretized random field and corresonds to the number of terms employed in \cite{Uribe2020}). The  Karhunen-Lo\`eve-expansion reads
\begin{equation}
	\label{eq:KL_E}
	E(x,y) = \exp\left\{\mu_{\ln E} + \sigma_{\ln E} \sum_{i=1}^{d_E} \sqrt{\lambda^E_i} \varphi^E_i(x,y) \xi_i\right\}.
\end{equation}
$\mu_{\ln E}$ and $\sigma_{\ln E}$ are the parameters of the log-normal marginal distribution of $E$, $\{\lambda^q_i,\varphi^E_i\}$ are the eigenpairs of the correlation kernel in $\cref{eq:corr_E}$ (see \cref{fig:plate}) and $\bm{\xi} \in \mathbb{R}^{d_E\times1}$ is a standard-normal random vector.
The most influential eigenfunctions (based on a global output-oriented sensitivity analysis of the plate model performed in \cite{Ehre2020b}) are shown in \cref{fig:plate} on the right.
\\~\\
Assuming plane stress, the displacement field  $\bm{u}(x_1,x_2)=[u_{1}(x_1,x_2),u_{2}(x_1,x_2)]^\mathrm{T}$ and the strain field 
$\bm{\varepsilon} = 0.5[\nabla \bm{u} + (\nabla \bm{u})^\mathrm{T}]$ can be computed implicitly based on elasticity theory through a set of elliptic partial differential equations (Cauchy-Navier equations) \cite{johnson_2009}:
\begin{equation}
	\nabla\left[ \frac{\nu E(x_1,x_2)}{1-\nu^2} \mathrm{tr}(\bm{\varepsilon})\mathbf{I} +  G(x_1,x_2) \bm{\varepsilon} \right] + \bm{B}  = 0.
	\label{eq:ellPDE}
\end{equation}
$G(x_1,x_2):=E(x_1,x_2)/(2(1+\nu))$ is the plate material's shear modulus, $E(x_1,x_2)$ is its Young's modulus, and $\bm{B}=[b(x_1),b(x_2)]^\mathrm{T}$ is the vector of (gravitational) body forces acting on the plate. In order to solve \cref{eq:ellPDE}, a finite element model with $282$ eight-noded quadrilateral finite elements is used (\cref{fig:plate}).
\\~\\
The LSF is defined by means of a threshold for the the first principal plane stress 
$$\sigma_1 = 0.5(\sigma_x + \sigma_y) + \sqrt{[0.5(\sigma_x + \sigma_y)]^2 + \tau_{xy}^2}$$ evaluated
at node 11 (see green marker \cref{fig:plate}, left).
Node 11 indicates a location where maximum plane stresses occur frequently in this example.
The LSF reads
\begin{equation}
	g(\bm{X}) = \sigma_{\mathrm{threshold}} - \sigma_1(\bm{X}),
\end{equation}
where $\bm{X} \vcentcolon=[q,\bm{\xi}]$ (and hence a problem dimension of $d = d_E + 1$
) and $\sigma_{\mathrm{threshold}} = 320$~MPa. The target probability of failure is determined as $p = 2.968 \cdot 10^{-6}$ with $\mathrm{CoV} = 0.0182$ as the average of 100 repeated runs of subset simulation \cite{Au2001} with $5 \cdot 10^3$ samples per level.
\paragraph{Results}
\cref{tab:plate_results} shows a comparison of SVRE using $n=1000$ and $n=5000$ estimation samples with iCEred \cite{Uribe2020} in terms of relative RMSE and the number of LSF and LSF gradient evaluations. iCEred is a subspace cross-entropy importance sampling method for rare event estimation that uses LSF gradients to inform the choice of subspace.
We run SVRE with $\ell_2$ normalization, base learning rate $\bar{\epsilon} = 1$ and $\delta_{\mathrm{thresh}} = 5$.
\cite{Uribe2020} report estimator coefficient of variation and mean calculated from 40 repeated runs, from which the relative RMSE is computed according to \cref{eq:rRMSE}.
\begin{table*}[h!]
	\setlength{\tabcolsep}{0.1cm}
	\centering
	\caption{Comparison or relative root mean squared error (rRMSE) and required number of model and model gradient evaluations of $p_{\mathcal{F}}$-estimates by CEBUred for rare events  \cite{Uribe2020} and our method with $n_\nabla =20$, $\delta_{\mathrm{thresh}} = 5$, $\bar{\epsilon} = 1$ and $\ell_2$ normalization.}
\label{tab:plate_results}
\begin{tabular}{c|c|ccc|ccc}
	\hline
	&& \multicolumn{3}{c|} {rRMSE} & \multicolumn{3}{c} {gradient (+ model) calls}  \\ 
	example  & dimension &  iCEred & SVRE &  SVRE   &  iCEred &SVRE  &  SVRE   \\
& &  &   ($n=1000$) &    ($n=5000$) &  &  ($n=1000$) &    ($n=5000$)  \\
	\hline
	\multirow{2}{*}{\textbf{Steel Plate}} &$d=869$ & \multirow{2}{*}{0.060} & \multirow{2}{*}{0.078} & \multirow{2}{*}{0.035} & 3037 &  98.0& 98.4  \\
	 &($p_{\mathcal{F}} = 2.968 \cdot 10^{-6}$) & & &  & (+3685) &   (+1000)& (+5000) \\
	\hline
\end{tabular}
\end{table*}
With 7.8 \%, SVRE using $1000$ estimation samples has a marginally larger relative RMSE than iCEred while  SVRE with $5000$ reduces the relative RMSE to 3.5\% while using approximately 3\% of the LSF gradient evaluations required for iCEred. However, the total number of iCEred gradient samples could likely be reduced without increasing relative RMSE by selecting $n_{\nabla}$ adaptively in each step. This has been introduced in the context of reduced space cross-entropy importance sampling for Bayesian inference in \cite{Ehre2023}. 
\begin{table}[H]
	\setlength{\tabcolsep}{0.1cm}
	\centering
	\caption{Comparison or relative root mean squared error (rRMSE) and required number of model and model gradient evaluations of $p_{\mathcal{F}}$-estimates by our method for different model dimensions with $n_\nabla =20$, $\delta_{\mathrm{thresh}} = 5$, $\bar{\epsilon} = 1$ and $n = 1000$.}
\label{tab:plate_results_dim}
\begin{tabular}{c|c|c|c}
	\hline
	example &dimension (average $p_{\mathcal{F}}$ from 500 SVRE runs ) &rRMSE  & gradient (+ model) calls  \\ 
	\hline
	 \multirow{3}{*}{\textbf{Steel Plate}}  &$d=100$ ($p_{\mathcal{F}} = 1.400 \cdot 10 ^{-6}$) & 0.111 & 86.8 (+$10^3$)   \\
	&$d=500$ ($p_{\mathcal{F}} =2.736 \cdot 10 ^{-6}$) & 0.097 & 84.7 (+$10^3$)   \\
	&$d=869$ ($p_{\mathcal{F}} =2.968 \cdot 10^{-6}$) & 0.078 & 98.0 (+$10^3$)   \\
	&$d=1500$ ($p_{\mathcal{F}} =2.968 \cdot 10^{-6}$) & 0.086 & 99.2 (+$10^3$)   \\
	\hline
\end{tabular}
\end{table}
\cref{tab:plate_results_dim} shows SVRE results in different dimensions for the plate problem. As in the diffusion problem, different $d$-settings correspond to the number of terms in the KLE. 
With $n=1000$ estimation samples the relative RMSE is approximately constant around 10\% in all dimensional settings. Likewise, the computational cost in terms of total gradient evaluations required seems insensitivity to $d$. The failure probability estimate decreases with decreasing dimension, which can be attributed to resolving less total variance of the Young's modulus random field with less KLE terms \cite{Uribe2020}. However, adding more terms to the Karhunen-Lo\`eve-expansion will not increase the failure probability estimate indefinitely and in particular between $d=869$ and $d=1500$, there is no significant change in target failure probability anymore. 
\section{Conclusion}
\label{sec:conclusion}
We have proposed a method termed Stein variational rare event simulation (SVRE) to sample from rare events and estimate the associated rare event probability. SVRE relies on Stein variational gradient descent (SVGD) -- an approach for sampling from non-normalized distributions proposed in the context of Bayesian inference of probabilistic machine learning models \cite{LiuWang2016}. We reformulate SVGD to move samples initially drawn from a tractable input distribution towards a near-optimal rare event importance sampling density. This is done by interpreting the indicator function of the rare event as a likelihood and smoothening it. Propagating samples repeatedly along streamlines of a specific set of velocity fields that are constructed in a reproducing kernel Hilbert space (RKHS) optimally reduces the KL divergence between the empirical distribution of the samples and the target distribution, i.e., the optimal rare event IS distribution of the smoothened failure event. Once the propagated samples have an empirical distribution sufficiently close to the targert (in terms of Kullback-Leibler-divergence), they are used to construct an importance sampling estimate of the rare event probability.
\\~\\
We investigate two different methods for rescaling the optimal velocity field in each step and suggest an approximation for tracking velocity field determinants that allows to track the density of the particles. The approximation facilitates adaptively choosing the base learning rate that controls the convergence speed of the method. Further we suggest an exit criterion for when to stop evolving samples and building the importance sampling estimate.
\\~\\
We gauge the performance of SVRE against three state-of-the-art gradient-based rare event sampling algorithms. We examine accuracy and efficiency of the SVRE estimator using five numerical examples spanning problem dimensions between $d=2$ and $d=1500$. We find our SVRE estimator performs better or on par with several state-of-the-art rare event simulation methods while reducing the required number of model gradient evaluations by at least one and in some cases up to three orders of magnitude. This is relevant as gradient evaluations both are responsible for most of the computational effort associated with gradient-based rare event simulation algorithms as well as the facilitator for superior performance in complicated problems and for extremely small target probabilities ($10^{-9}$ -- $10^{-12}$) compared to gradient-free methods.
\\~\\ 
While we show that the SVRE estimator works well in several high-dimensional problems, the underlying SVGD method has been found to suffer from a variance degeneracy in high-dimensions \cite{Ba2022}. Using distance-based kernels presents a second degeneracy that likely needs to be addressed to ensure SVRE works well in a very general high-dimensional setting \cite{Ramdas2015}. An interesting direction for future research may thus be to flexibilize or generalize SVRE to work well in generic high dimensions, for example by investigating non-distance-based and heterotropic kernels. Alternatively, it is straightforward to combine the SVRE with subspace approaches proposed in the context of SVGD such as \cite{Chen2020,Liu2022}.
\\~\\
Given that the number of gradient evaluations is the computational bottleneck in the rare event setting, locally replacing $g$ with a cheap gradient-true surrogate model may facilitate further efficiency gains. Such a surrogate model can be integrated in SVRE by considering a Stein operator with respect to the surrogate model and reweighting the RKHS by the likelihood ratio of the surrogate and the true model \cite{Han2018}.
\section{Acknowledgment}
This project was supported by the German Research Foundation (DFG) through grants STR 1140/11-1 and PA2901/1-1.
\newpage
\appendix
\section{Jacobian of the particle transformation at step $t$}
\label{appendix:B}
The particle transformation in the $t$-th step of SVGD reads
\begin{align*}
	\mathcal{T}_t(\bm{x}) &= \bm{x} + \epsilon_t(\bm{x}) \odot  \phi_t(\bm{x})\\
	\bm{\phi}_t(\bm{x}) &= \sum\limits_{i=1}^{n_{\nabla}} k(\bm{x},\bm{y}_i) \nabla_{\bm{y}} \ln p(\bm{y}_i) + \nabla_{\bm{y}} k(\bm{x},\bm{y}_i).
\end{align*}
This transformation has the form of a gradient descent with $\bm{\phi}_t(\bm{x})$ representing the objective gradient at $\bm{x}$ and $\epsilon_t$ is the step width or learning rate. $\epsilon_t(\bm{x})$ may be adjusted along each coordinate of $\mathcal{X}$ individually.
In the stochastic gradient descent literature, it is common practice to rescale $\epsilon_t(\bm{x})$ based on past and current gradient fields, hence it depends on $\bm{x}$ through $\bm{\phi}_t$. The Jacobian of $\mathcal{T}_t(\bm{x})$ reads
\begin{equation*}
\nabla_{\bm{x}} \mathcal{T}_t(\bm{x}) = \mathbf{I} +  \diag \epsilon_t(\bm{x}) \cdot \nabla_{\bm{x}} \phi_t(\bm{x}) + \diag \phi_t(\bm{x}) \cdot \nabla_{\bm{x}}  \epsilon_t(\bm{x}).
\end{equation*}
where
\begin{equation*}
	\nabla_{\bm{x}}  \phi_t(\bm{x}) = \sum\limits_{i=1}^{n_{\nabla}} \nabla_{\bm{x}} k(\bm{x},\bm{y}_i) \nabla_{\bm{y}} \ln p(\bm{y}_i) + \nabla_{\bm{x}} \nabla_{\bm{y}} k(\bm{x},\bm{y}_i).
\end{equation*}
\\~\\
In the original SVGD paper, the authors suggest using Adagrad \cite{Duchi2011} (or more precisely, a slightly improved version thereof called RMSProp) to select $\epsilon_t$. RMSProp rescales the current gradient field with an exponential moving average of past gradient fields, which reads 
\begin{align*}
	\epsilon_t(\bm{x}) &= \frac{\bar{\epsilon}_t}{\varepsilon + v_t(\bm{x})} \\
	  v_t^2(\bm{x}) &= \begin{cases} \phi_0^2,& t = 0 \\ \alpha^t\phi_0^2 + (1 - \alpha) \sum_{t'=1}^{t} \alpha^{t-t'}\phi_{t'}^2, & t > 0 \end{cases}.
\end{align*}
Therein, $\bm{\phi}_{t'}$ is the gradient field in step $t' \leq t$ and $\bar{\epsilon}_t$ is a base learning rate that acts equally on all coordinates and particles. The velocity of the moving average is controlled by $\alpha$, which is usually selected to be $\alpha = 0.9$. $\varepsilon = 10^{-6}$ is a small nugget to avoid divide-by-zero errors. 
\\~\\
Ignoring the nugget, the Jacobian of the RMSProp learning rate is $\nabla_{\bm{x}} \epsilon_t (\bm{x}) \approx - \bar{\epsilon}_t  \diag v_t^{-2} (\bm{x}) \cdot \nabla_{\bm{x}} v_t (\bm{x})$, where $\bm{a}^{b}$ with $\bm{a}$ a vector and $b$ an integer denotes an elementwise power.
We define 
\begin{equation*}
	\bm{\phi}_{\text{old}}^2 \vcentcolon=  \alpha^t\phi_0^2 + (1 - \alpha) \sum_{t'=1}^{t-1} \alpha^{t-t'}\phi_{t'}^2
\end{equation*}
and obtain
\begin{equation*}
	\nabla_{\bm{x}}  v_t(\bm{x})  = \begin{cases} \nabla_{\bm{x}} \phi_t(\bm{x}), &  t = 0 \\  (1- \alpha)  \diag (\phi_t(\bm{x}) / v_t(\bm{x}))  \cdot \nabla_{\bm{x}} \phi_t(\bm{x}), & t >0 \end{cases},
\end{equation*}
where $\bm{a} / \bm{c}$ with $\bm{a}$ and $\bm{c}$ vectors is an elementwise division.
Hence,
\begin{equation*}
	\nabla_{\bm{x}}  \mathcal{T}_t(\bm{x}) 
	\approx 
	\mathbf{I} +  \begin{cases} 0, &  t = 0 \\  \bar{\epsilon}_t \diag (\phi_{\text{old}}^2(\bm{x}) / v_t^3(\bm{x}))  \cdot \nabla_{\bm{x}} \phi_t(\bm{x}), & t >0 \end{cases}
\end{equation*}
That is,  $\phi_{\text{old}}^2/ v_t^{3}(\bm{x})$ adjusts the Jacobian of $\mathcal{T}_t$ for the adapative learning rate selection performed in RMSProp. 
\\~\\
Instead of adjusting the learning rate differently along each dimension of $\mathcal{X}$ as in RMSProp, we may want to preserve the original gradient field, so that $\epsilon_t$ is a scalar depending on $\bm{x}$. In this case
\begin{align*}
	\mathcal{T}_t(\bm{x}) &= \bm{x} + \epsilon_t(\bm{x})  \phi_t(\bm{x})\\
	\bm{\phi}_t(\bm{x}) &= \sum\limits_{i=1}^{n_{\nabla}} k(\bm{x},\bm{y}_i) \nabla_{\bm{y}} \ln p(\bm{y}_i) + \nabla_{\bm{y}} k(\bm{x},\bm{y}_i).
\end{align*}
The Jacobian of  $\mathcal{T}_t$ in this case reads
\begin{equation*}
	\nabla_{\bm{x}} \mathcal{T}_t(\bm{x}) = \mathbf{I} +  \epsilon_t(\bm{x}) \cdot \nabla_{\bm{x}} \phi_t(\bm{x}) +  \phi_t^T(\bm{x}) \cdot \nabla_{\bm{x}}  \epsilon_t(\bm{x}).
\end{equation*}
An intuitive way to choose $\epsilon_t$ is to normalize each gradient evaluation, i.e., set $\epsilon_t(\bm{x}) =  \bar{\epsilon}_t \lVert \phi_t(\bm{x}) \rVert^{-1}$. With 
\begin{equation*}
	\nabla_{\bm{x}}  \lVert \phi_t(\bm{x}) \rVert^{-1} = - \frac{\nabla_{\bm{x}}  \lVert \phi_t(\bm{x}) \rVert}{ \lVert \phi_t(\bm{x}) \rVert^2} = -  \nabla_{\bm{x}}  \phi_t(\bm{x}) \cdot \phi_t(\bm{x}) / \lVert \phi_t(\bm{x}) \rVert^3
\end{equation*}
we have
\begin{equation}
	\nabla_{\bm{x}}  \mathcal{T}_t(\bm{x}) \approx \mathbf{I} + \bar{\epsilon_t} \frac{\nabla_{\bm{x}}  \phi_t(\bm{x})}{\rVert  \phi_t(\bm{x})  \lVert } \cdot \left(\mathbf{I} -  \frac{\phi_t^\T(\bm{x}) \cdot \phi_t(\bm{x}) }{ \rVert  \phi_t(\bm{x})  \lVert^2}\right).
\end{equation}
\section{Trace approximation of $\mathrm{det} (\nabla_{\mathbf{x}}  \mathcal{T}_t)$}
\label{appendix:trace_approx}
Evaluating log PDF of particles in generation $t$,
\begin{equation*}
	 \log q_\ell (\bm{x}) = \log q_0 (\bm{x}) - \sum_{i=1}^t \log | \mathrm{det}(\nabla_{\bm{x}} \mathcal{T}_t(\bm{x}))|,
\end{equation*}
requires evaluating $ \mathrm{det}(\nabla_{\bm{x}}  \mathcal{T}_t(\bm{x}))$ at all $n$ particle locations, which costs $\mathcal{O}(nd^3)$. For small $\bar{\epsilon}_t$,
 \begin{equation*}
 	 \mathrm{det} \nabla_{\bm{x}}  \mathcal{T}_t(\bm{x}) =  \mathrm{det}(\mathbf{I} + \bar{\epsilon}_t  \mathbf{A}) \approx   1 + \bar{\epsilon}_t \mathrm{trace}(\mathbf{A}) ~~(+ \mathcal{O}(\bar{\epsilon}_t^2))
 \end{equation*}
is a useful approximation that reduces the overall cost from $\mathcal{O}(nd^3)$ to $\mathcal{O}(nd)$.
Therein,
 \begin{equation*}
A_{ij} = \epsilon_t(\bm{x}_i)\nabla_{x_j} \phi_t(\bm{x}_i) + \phi_t(\bm{x}_i) \nabla_{x_j}  \epsilon_t(\bm{x}_i)  \text{~($\ell_2$)} \text{~~~or~~~} A_{ij} = \alpha  \frac{ \nabla_{x_j} \phi_t(\bm{x}) \phi_{\text{old},ij}^2 }{v_{t,ij}^{3}(\bm{x})} \text{~(RMSProp)}.
 \end{equation*}
 Further, the approximation facilitates explicitly choosing $\bar{\epsilon}_t$ such that $\mathrm{det}(\nabla_{\bm{x}} \mathcal{T}_t(\bm{x}))$ is bounded at all particle locations and hence $\mathcal{T}_t(\bm{x})$ is locally invertible at all particle locations.
\bibliographystyle{siamplain}
\bibliography{references}

\end{document}